\documentclass[reprint, amsmath,amssymb, aps,prd, groupedaddress, nofootinbib]{revtex4}
\usepackage{graphicx}
\usepackage{bm}
\usepackage{tikz}
\usetikzlibrary{arrows.meta,calc}
\usepackage{hyperref}
\hypersetup{
citecolor=red,
colorlinks=true,
filecolor=red,
linkcolor=darkgreen,
linktocpage=true,
urlcolor=darkblue
}
\definecolor{darkgreen}{rgb}{0,0.5,0}
\definecolor{darkblue}{rgb}{0,0,0.6}

\newcommand{\Bbar}{\bar{B}}
\newcommand{\SR}{S_{R}}
\newcommand{\tr}{\operatorname{Tr}}
\newcommand{\ket}[1]{|#1\rangle}
\newcommand{\bra}[1]{\langle #1|}
\newcommand{\ketbra}[2]{|#1\rangle\langle #2|}
\newcommand{\nn}{\bar{n}}

\begin{document}

\title{Surviving correlations across a horizon: reflected entropy for bosonic fields in non-inertial frames and black hole spacetimes}

\author{Sayid Mondal}
\email[]{sayid.mondal@gmail.com}
\affiliation{Instituto de Ciencias Exactas y Naturales,\\ Universidad Arturo Prat, Playa Brava 3256, 1111346, Iquique, Chile}

\begin{abstract}
We study the reflected entropy and the Markov gap for modes of a free bosonic
field shared by inertial observers (Alice, Charlie) and a uniformly accelerated
one (Bob), for a bipartite Bell state and the tripartite Werner (W) and
Greenberger--Horne--Zeilinger (GHZ) states. The bosonic Bogoliubov transformation spans an infinite-dimensional Fock space with an unbounded squeezing parameter unlike the fermionic case. By identifying a conserved charge, we block-diagonalize the reduced density matrices into exact two-dimensional sectors, yielding closed or semi-analytic forms
for all three states. Although bosonic entanglement is known to vanish asymptotically, the Alice--Bob reflected entropy instead saturates at a nonzero floor, retaining the surviving classical correlation, and
converges to the value Alice shares with Bob's causally disconnected partner.
 Crucially, only the inter-wedge reflected entropy diverges, linearly in the squeezing
parameter---the sharp distinction from the fermionic case, where it stays bounded---while
the Markov gaps saturate. The construction transfers verbatim to a Schwarzschild
black hole, where the saturation values become mass-independent constants.

\end{abstract}

\maketitle
\tableofcontents
\section{Introduction}
\label{sec:intro}

One of the more striking lessons of combining quantum information with relativity
is that entanglement is \emph{observer dependent}. Two inertial observers who
share a maximally entangled pair agree on how entangled it is; but if one of them
accelerates, that observer no longer sees the same vacuum. A uniformly
accelerated detector responds to the inertial vacuum as though immersed in a
thermal bath at the Unruh temperature $T=a/2\pi$ \cite{Unruh1976,Davies1975}.
The thermal noise this represents degrades the correlations the accelerated
observer can access. Fuentes-Schuller and Mann first quantified this for a scalar
field \cite{FuentesSchuller2005}, and Alsing \textit{et al.} for Dirac fields
\cite{Alsing2006}. A robust qualitative split emerged between the two statistics:
fermionic entanglement degrades to a nonzero minimum at infinite acceleration,
whereas bosonic entanglement can be destroyed completely
\cite{Adesso2007,FuentesSchuller2005}. The continuous-variable analysis of
Adesso \textit{et al.} \cite{Adesso2007} traced this to the
unbounded occupation of bosonic Rindler modes and reframed the loss as a
\emph{redistribution} of correlations among the accessible and inaccessible
regions of spacetime. The attribution deserves care: the authors in \cite{MartinMartinezLeon2010b} argued that the decisive factor is the field statistics itself rather than the mere dimensionality of the single-mode Fock space. Tripartite versions followed for both statistics
\cite{NasrEsfahani2011,Hwang2011,TorresArenas2019}, and the validity of the single-mode
approximation underlying these treatments was examined in \cite{Bruschi2010}.

Almost all of these works quantify correlations with the entanglement entropy, the
negativity, or various tangles. For a \emph{mixed} state, however, the
entanglement entropy is not a faithful correlation measure---and the states an
accelerated observer sees are generically mixed, because the second Rindler wedge
is inaccessible and must be traced out. A measure better suited to mixed states
is the \emph{reflected entropy}, proposed by Dutta and Faulkner
\cite{DuttaFaulkner2021}. The idea is to restore purity before
measuring: one adjoins a mirror copy of the system, builds a canonical
purification in which the mixedness is recorded in the mirror copy, and then measures
how entangled a party is with its own mirror image. Correlations that were
present in the mixed state are ``reflected'' into entanglement in the doubled
system, whence the name. The construction is natural rather than ad hoc---the
purification is canonical, fixed by the state itself---and holographically the
resulting quantity is dual to (twice) the entanglement-wedge cross section
\cite{TakayanagiUmemoto2017,Nguyen2017,DuttaFaulkner2021}, which is what has made
it a standard tool in the gravitational literature. The excess of reflected entropy over the mutual information defines the {\it Markov gap} introduced by Hayden {\it et al.} in \cite{HaydenParrikarSorce2021}. It equals a conditional mutual information in
the purified state, and by the Fawzi--Renner theorem \cite{FawziRenner2015} it
controls how well the global state can be reconstructed by an operation performed
on one party alone: a large gap certifies that the correlation is irreducibly
tripartite, invisible to any bipartite measure. Together, then, the pair
(reflected entropy, Markov gap) answers two questions the entanglement entropy
cannot: how much correlation of any kind survives, and how much of it is genuinely
shared among three parties. 

One caveat must be stated at the outset: reflected entropy is not monotonic under
partial trace---discarding part of one party can \emph{increase} it
\cite{HaydenLemmSorce2023}---so it is not a correlation measure in the axiomatic
sense. None of our conclusions rests on that property. The failure concerns
coarse-graining, whereas our partition is fixed by causal structure: Alice and
Charlie hold single qubits, Bob and his partner single modes across a horizon, so
no sub-party is available to discard and we follow one fixed bipartition as the
acceleration varies. What we use is faithfulness---reflected entropy vanishes
only on product states and dominates the mutual information---so a nonvanishing
limit certifies surviving correlation; and the mutual information, a genuine
monotone, saturates likewise, showing the effect to
be a property of the state rather than of the diagnostic. It is also worth noting
that the known violations are exceptional rather than generic: in explicit
free-field computations, Bueno and Casini found the conjectured monotonicity
under enlargement of one party to hold for both free scalars and free fermions
\cite{BuenoCasini2020scalars,BuenoCasini2020fermions}---precisely the class of
theories considered here. Taken together, these considerations justify
the reflected entropy as our probe of bipartite correlation, and the Markov gap
as its tripartite companion.

The degradation of field-mode correlations under acceleration has been studied
extensively, but almost exclusively with entanglement-type measures. For bosonic
fields, Fuentes {\it et al.} showed that a maximally entangled two-mode
state loses its distillable entanglement entirely in the infinite-acceleration
limit \cite{FuentesSchuller2005}, in contrast to the fermionic case, where the
entanglement saturates at a nonzero saturation floor \cite{Alsing2006}. This species
dichotomy was extended to tripartite GHZ and W states by Nasr {\it et al.} \cite{NasrEsfahani2011}, who quantified it with negativity
and tangles, later revisited for the W state with a range of entanglement
monotones \cite{TorresArenas2019}, to the case of two accelerated parties
\cite{RichterOmar2015}, and to other correlation measures such as
measurement-induced nonlocality \cite{TianJing2013}; in every case bosonic correlations decay to zero
asymptotically while fermionic ones do not.

An important exception to this pattern was found early, with \emph{quantum
discord}. Datta computed the discord between two relatively accelerated
scalar-field modes and showed that a finite amount of purely quantum correlation
survives even where distillable entanglement is gone, with evidence for a nonzero
residue at infinite acceleration \cite{Datta2009}. Companion studies mapped the
distribution of classical correlation and discord across the wedges for Dirac
fields \cite{WangDengJing2010} and, comparing statistics near a horizon, found
that for the scalar field the classical correlation is essentially insensitive to
the acceleration while the entanglement degrades \cite{MartinMartinezLeon2010}---a
fact we invoke below to interpret our saturation results. A similar saturation has also been reported for mutual information and quantum coherence, both in non-inertial frames and in black-hole backgrounds
\cite{Pan:2008yr,WuZengCao2021,LiWu2024,Teng:2026dyr}; see \cite{Alsing:2012wf} for a review of the earlier literature.

In parallel, the reflected entropy \cite{DuttaFaulkner2021} and the Markov gap
\cite{HaydenParrikarSorce2021} have emerged as sharper diagnostics of mixed-state
and multipartite correlation. They have since been evaluated in a range of
settings---free scalar and fermion theories
\cite{BuenoCasini2020scalars,BuenoCasini2020fermions,DuttaFaulknerLin2023,BerthiereParez2023},
conformal field theory \cite{Camargo2021}, Lifshitz theories
\cite{BerthiereChenChen2023}, random tensor networks
\cite{AkersFaulknerLinRath2022}, and evaporating black holes
\cite{LiChuZhou2020,AkersPageCurve2022,LuLin2023}---and, in the non-inertial
setting, for \emph{fermionic} modes \cite{Basak2023}. What has not been done is to bring these
mixed-state measures to bear on \emph{bosonic} modes in non-inertial frames. We
do so here, and find that reflected entropy refines the established entanglement
picture in the same spirit as the discord results, but more sharply: although the
bipartite entanglement vanishes, the Alice--Bob reflected entropy
\emph{saturates} at a finite floor, because it retains the surviving classical
correlation. The genuinely unbounded effect is confined to the inter-wedge
pair, which has no fermionic analogue, where all quantities remain bounded.

Passing from fermions to bosons is not a matter of changing a sign. For fermions
the Pauli exclusion principle truncates each mode to a two-level system, so every density
matrix is small and finite, and the fermionic squeezing parameter is confined to
a bounded interval. For bosons neither simplification holds: a single inertial
excitation spreads over the entire Fock tower of each wedge, the mean occupation
of the accelerated mode grows without bound, and the squeezing parameter sweeps
the whole half-line. Even keeping Alice and Charlie as qubits, Bob's accelerated
mode is genuinely infinite dimensional, and no closed two-level density matrix
exists. The technical content of this paper is a set of exact methods that make
this infinite problem finite: a reduction of the Alice-pairs to a
four-dimensional matrix that is exact at any cutoff, a closed form for the GHZ
state, a decomposition of the Bell and W states into an infinite family of
two-dimensional blocks labelled by a conserved charge, and an exact identity with
a hidden block structure for the inter-wedge pair. 

The central physical results, valid over the full range of acceleration, are the
following. First, the Alice--Bob reflected entropy degrades as the acceleration
grows but \emph{saturates} at a finite nonzero floor rather than decaying to
zero; for the GHZ state that floor is an exact constant, and for the maximally
entangled Bell state it is given by an exponential integral. Both are pure
numbers, independent of the details of the state's preparation. Second, the
reflected entropies of Bob and of his causally disconnected partner tend to a
\emph{common} limit, which is the bosonic expression of the two wedges becoming
statistically indistinguishable at infinite acceleration. Third, the only
quantity that grows without bound is the inter-wedge reflected entropy, which
diverges linearly in the squeezing parameter, driven by the unbounded thermal
occupation of bosonic modes; this, rather than the behaviour of the Alice--Bob
pair, is the sharp fermion/boson discriminator, since the corresponding
 quantity for the fermionic case is saturated to a finite value (see \cite{Basak2023}). Fourth, the Markov gaps saturate and reproduce the state dependence found for fermions, indicating that the residual
correlation retains a genuinely tripartite component. Finally, because these
results depend on the state only through the squeezing parameter, they transfer
intact to a Schwarzschild black hole, where the same saturation floors reappear as
mass-independent constants.

The paper is organized as follows. Section~\ref{sec:setup} sets up the field
theory, derives the two-mode squeezed structure of the accelerated vacuum and its
thermofield-double interpretation, and introduces the three states we study.
Section~\ref{sec:reflected} defines the canonical purification and the reflected
entropy, and establishes the finite reduction that makes the
infinite-dimensional problem tractable; this machinery is then applied to the GHZ
state (Sec.~\ref{sec:ghz}), the Bell state (Sec.~\ref{sec:bellworked}) and the
W state (Sec.~\ref{sec:wstate}), with the resulting reflected entropies
discussed in Sec.~\ref{sec:results}. Section~\ref{sec:markov} defines the Markov
gap and analyzes the corresponding results for the same three states. Section~\ref{sec:blackhole}
extends the analysis to a Schwarzschild black hole spacetime.
Finally, Section~\ref{sec:discussion} summarizes our findings and lists open directions,
and an appendix collects the explicit reduced density matrices.

\section{Setup: bosonic modes and the states}
\label{sec:setup}

\subsection{The accelerated mode}

\begin{figure}[t]
\centering
\begin{tikzpicture}[scale=1.35,>=Latex]
  \draw[thick,gray] (-2,-2) -- (2,2);
  \draw[thick,gray] (-2,2) -- (2,-2);
  \fill[blue!12] (0,0) -- (2,2) -- (2,-2) -- cycle;
  \fill[red!10]  (0,0) -- (-2,2) -- (-2,-2) -- cycle;
  \fill[gray!8]  (0,0) -- (2,2) -- (-2,2) -- cycle;
  \fill[gray!8]  (0,0) -- (2,-2) -- (-2,-2) -- cycle;
  \draw[->,thick] (-2.2,0) -- (2.25,0) node[right] {$x$};
  \draw[->,thick] (0,-2.05) -- (0,2.1) node[above] {$t$};
  \foreach \a in {0.7,1.05}{
    \draw[blue!60!black,thick,domain=-1.2:1.2,smooth,variable=\u]
      plot ({\a*cosh(\u)},{\a*sinh(\u)});}
  \draw[->,blue!60!black,very thick,domain=-1.0:1.0,smooth,variable=\u]
      plot ({1.05*cosh(\u)},{1.05*sinh(\u)});
  \foreach \a in {0.7,1.05}{
    \draw[red!60!black,thick,domain=-1.2:1.2,smooth,variable=\u]
      plot ({-\a*cosh(\u)},{\a*sinh(\u)});}
  \draw[->,black,very thick] (0.3,-1.75) -- (0.3,1.95) node[above right]{\footnotesize $A$};
  \node[gray!70!black] at (1.45,1.78) {\footnotesize $\mathcal{H}$};
  \node[blue!55!black] at (1.75,0.2) {\textbf{I}};
  \node[red!55!black]  at (-1.75,0.2) {\textbf{II}};
  \node[blue!55!black,align=center] at (1.32,-0.7)
     {\footnotesize ~~~$B$};
  \node[red!55!black,align=center] at (-1.3,-0.7)
     {\footnotesize $\Bbar$~~~};
  \node[gray!55!black] at (0.16,1.62) {\footnotesize F};
  \node[gray!55!black] at (0.16,-1.62) {\footnotesize P};
\end{tikzpicture}
\caption{Minkowski spacetime in Rindler coordinates. The lightlike lines
$x=\pm t$ form the Rindler horizon $\mathcal{H}$, dividing space into the right
(I) and left (II) wedges. A uniformly accelerated observer, Bob ($B$), follows a
hyperbola $x^2-t^2=\mathrm{const}$ (blue) confined to region~I; the causally
disconnected region~II defines anti-Bob ($\Bbar$). The observer Alice ($A$) is inertial. P and F denote the past and future Rindler wedges respectively.}
\label{fig:wedge}
\end{figure}
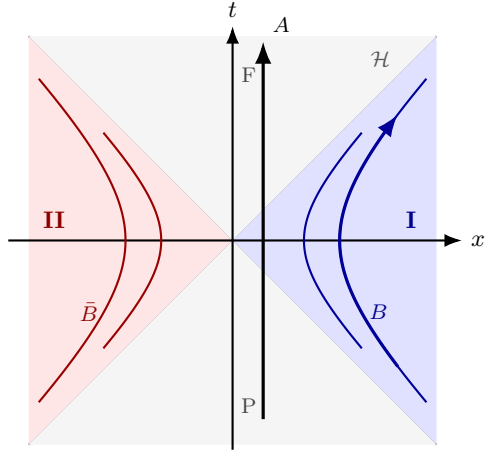

A uniformly accelerated observer is described in Rindler coordinates, which cover
two causally disconnected wedges of Minkowski space: region~I, whose observer we
call Bob ($B$), and region~II, anti-Bob ($\Bbar$) (Fig.~\ref{fig:wedge}).
Quantising a free massless
scalar in these coordinates and relating Minkowski operators to Rindler ones by a
Bogoliubov transformation, Bob's Minkowski vacuum becomes a two-mode squeezed
state shared across the wedges \cite{Adesso2007,FuentesSchuller2005}

\begin{equation}
\ket{0}_M=\frac{1}{\cosh r}\sum_{n=0}^{\infty}\tanh^{n}\! r\,\ket{n}_B\ket{n}_{\Bbar},
\label{eq:vac}
\end{equation}
where $\ket{n}_B$ and $\ket{n}_{\Bbar}$ are the two modes associated to the observer $B$ and $\bar B$. A single Minkowski excited state $\ket{1}_M=a_M^{\dagger}\ket{0}_M$ is
\begin{equation}
\ket{1}_M=\frac{1}{\cosh^{2} r}\sum_{n=0}^{\infty}\sqrt{n+1}\,\tanh^{n}\! r\,
\ket{n+1}_B\ket{n}_{\Bbar}\;,
\label{eq:one}
\end{equation}
where $a_M^\dagger=\cosh r\,b_{B}^\dagger-\sinh r\,b_{\Bbar}$ with $(b_{B},b_{B}^\dagger)$ and $(b_{\Bbar},b_{\Bbar}^\dagger)$ are creation and annihilation operators for $B$ and $\bar B$ respectively. The acceleration enters only through the squeezing parameter $r$, fixed by the
mode frequency $\omega$ and proper acceleration $a$ through
\begin{equation}
\tanh r=e^{-\pi\omega/a},\qquad 0\le r<\infty .
\label{eq:rdef}
\end{equation}
The inertial limit is $r\to0$ while the infinite acceleration is $r\to\infty$. Note that, $\tanh r=e^{-\pi\omega/a}\to1$ as $a\to\infty$ for bosonic case and the parameter $r$ sweeps the \emph{entire} half-line $[0,\infty)$ unlike case for fermions
where $\tan r=e^{-\pi\omega/a}$, restricting $r$ at $\pi/4$ as a consequence of Pauli exclusion. 

Moreover, throughout this paper, as in
the fermionic analysis \cite{Basak2023}, we work in the \emph{single-mode approximation}: Bob's detector is
assumed to couple to a single Unruh frequency $\omega$, so that a single
Minkowski mode maps onto a single pair of Rindler modes
\eqref{eq:vac}--\eqref{eq:one}. However, Bruschi \textit{et al.}~\cite{Bruschi2010}
showed that for general states a physical detector instead responds to a
one-parameter family of Unruh modes, and quantitative corrections can arise. 

\subsection{The two-mode squeezed vacuum and its thermofield-double character}
\label{sec:tmsv}

The eq.~\eqref{eq:vac} is a \emph{two-mode squeezed vacuum} (TMSV), and
recognising it as such clarifies the thermal nature of the reduced state. The Bogoliubov transformation relating
Minkowski and Rindler operators is generated by the two-mode squeeze operator
$S(r)=\exp[r(b_B^\dagger b_{\Bbar}^\dagger-b_B b_{\Bbar})]$, under which
$S^\dagger b_B S=\cosh r\,b_B-\sinh r\,b_{\Bbar}^\dagger$. Therefore, eq.~\eqref{eq:vac} is obtained as $|0\rangle_M=S(r)|0\rangle_R$, where $|0\rangle_R=|0\rangle_B|0\rangle_{\bar B}$ is the Rindler vacuum (annihilated by $b_B, b_{\bar B}$). This is already a Schmidt
decomposition, with Schmidt coefficients $\lambda_n=\tanh^n r/\cosh r$ and
unbounded Schmidt rank---in contrast to the fermionic case, where Pauli exclusion
truncates the analogous sum at $n=1$.

Tracing out the causally inaccessible wedge yields Bob's marginal
\begin{equation}
\rho_B=\tr_{\Bbar}\ket{0}_M\!\bra{0}_M
=(1-t^2)\sum_{n\ge0}t^{2n}\ketbra{n}{n}=\tau_B \;.
\label{eq:thermalmarg}
\end{equation}
 Writing
$t^2=\tanh^2 r\equiv e^{-\beta\omega}$ and using $\tanh r=e^{-\pi\omega/a}$ fixes
$\beta=2\pi/a$, so \eqref{eq:thermalmarg} is a Bose--Einstein thermal state
$\rho_B=e^{-\beta\omega\, b_B^\dagger b_B}/Z$ at the \emph{Unruh temperature}
$T=1/\beta=a/2\pi$ \cite{Unruh1976,Alsing:2012wf,CrispinoHiguchiMatsas2008}, with partition
function $Z=\cosh^2 r$ and mean occupation 
\begin{equation}
\bar{n}=\operatorname{Tr}\left(\rho_B b_B^{\dagger} b_B\right)=\sinh^2 r\;.
\end{equation}

An observer confined to one Rindler wedge, with access to only half of the TMSV,
therefore registers a thermal bath---the algebraic content of the Unruh effect.

Equivalently, in the notation $\lambda_n=e^{-\beta E_n/2}/\sqrt{Z}$ with
$E_n=\omega n$, \eqref{eq:vac} is the \emph{thermofield double} (TFD) state
\begin{equation}
\ket{0}_M=\frac{1}{\sqrt Z}\sum_{n}e^{-\beta E_n/2}\ket{n}_B\ket{n}_{\Bbar},
\label{eq:tfd}
\end{equation}
the canonical purification of the Gibbs state $\rho_B=e^{-\beta H_B}/Z$ into the
doubled Hilbert space $\mathcal H_B\otimes\mathcal H_{\Bbar}$
\cite{TakahashiUmezawa1975,Israel1976}. This is the same canonical-purification
construction we apply to reflected entropy in Sec.~\ref{sec:reflected}, here
applied to the vacuum itself. The TFD is cyclic and separating for the wedge
algebra, and its Tomita--Takesaki modular Hamiltonian is
$K_B=-\log\rho_B=\beta H_B+\mathrm{const}$, so the reduced state obeys the KMS
condition at inverse temperature $\beta$ \cite{TakahashiUmezawa1975}. That the
modular flow of the Rindler wedge in the global Minkowski vacuum is precisely the
boost, with KMS temperature $a/2\pi$, is the content of the
Bisognano--Wichmann theorem \cite{BisognanoWichmann1975,Sewell1982}; the TMSV is
its single-mode avatar. In gravitational language the same structure is the
Hartle--Hawking state of an eternal (two-sided) black hole, dual across the
horizon to a thermofield double of two copies of the theory
\cite{Israel1976,Maldacena2003}---the two Rindler wedges $B$ and $\Bbar$ playing
the role of the two exterior regions.

\subsection{The three states}

In our setup, Alice ($A$) and Charlie ($C$) are inertial observers and Bob ($B$) accelerates. We consider three states, namely Bell, W and GHZ which are the bosonic analogues of those in \cite{Basak2023}
\begin{equation}
    \begin{aligned}
\ket{B}_{AB}&=\alpha\ket{0}_A\ket{0}_B+\sqrt{1-\alpha^{2}}\,\ket{1}_A\ket{1}_B,
\quad \alpha\in(0,1),\\
\ket{W}_{ABC}&=\alpha\ket{1}_A\ket{0}_B\ket{0}_C+\alpha\ket{0}_A\ket{0}_B\ket{1}_C+\sqrt{1-2\alpha^{2}}\,\ket{0}_A\ket{1}_B\ket{0}_C,
\quad \alpha\in\!\big(0,\tfrac{1}{\sqrt2}\big),\\
\ket{\mathrm{GHZ}}_{ABC}&=\alpha\ket{0}_A\ket{0}_B\ket{0}_C
+\sqrt{1-\alpha^{2}}\,\ket{1}_A\ket{1}_B\ket{1}_C.
\end{aligned}
\end{equation}
Substituting \eqref{eq:vac}--\eqref{eq:one} for Bob's modes promotes each to a
state shared by three parties  $A,B,\Bbar$ for Bell state, and 
four parties $A,B,\Bbar,C$ for W and GHZ states, which are given as follows
\begin{widetext}
\begin{align}
\ket{B}_{AB\Bbar}
&=\frac{\alpha}{\cosh r}\sum_{n=0}^{\infty}\tanh^{n}\!r\,
   \ket{0}_A\ket{n}_B\ket{n}_{\Bbar}
   +\frac{\sqrt{1-\alpha^{2}}}{\cosh^{2}r}\sum_{n=0}^{\infty}\sqrt{n+1}\,\tanh^{n}\!r\,
   \ket{1}_A\ket{n{+}1}_B\ket{n}_{\Bbar},
   \label{eq:bellsub}\\[4pt]
\ket{W}_{AB\Bbar C}
&=\frac{1}{\cosh r}\sum_{n=0}^{\infty}\tanh^{n}\!r\,
   \big(\alpha\ket{1}_A\ket{0}_C+\alpha\ket{0}_A\ket{1}_C\big)\ket{n}_B\ket{n}_{\Bbar}
   \nonumber\\
&\qquad
   +\frac{\sqrt{1-2\alpha^{2}}}{\cosh^{2}r}\sum_{n=0}^{\infty}\sqrt{n+1}\,\tanh^{n}\!r\,
   \ket{0}_A\ket{n{+}1}_B\ket{n}_{\Bbar}\ket{0}_C \;,
   \label{eq:wsub}\\[4pt]
\ket{\mathrm{GHZ}}_{AB\Bbar C}
&=\frac{\alpha}{\cosh r}\sum_{n=0}^{\infty}\tanh^{n}\!r\,
   \ket{0}_A\ket{n}_B\ket{n}_{\Bbar}\ket{0}_C
   +\frac{\sqrt{1-\alpha^{2}}}{\cosh^{2}r}\sum_{n=0}^{\infty}\sqrt{n+1}\,\tanh^{n}\!r\,
   \ket{1}_A\ket{n{+}1}_B\ket{n}_{\Bbar}\ket{1}_C\;.
   \label{eq:ghzsub}
\end{align}
\end{widetext}

At $r=0$ these reduce to the ordinary qubit states with $\Bbar$ a spectator
vacuum, a useful limit check. We study the three pairs $A{:}B$, $A{:}\Bbar$ and
$B{:}\Bbar$, tracing out the remaining party in each.

To obtain the reduced density matrix for the Bell state, we write \eqref{eq:bellsub} as
$\ket{B}=\alpha\ket{0}_A\ket{\psi_0}+\beta\ket{1}_A\ket{\psi_1}$ with
$\beta=\sqrt{1-\alpha^2}$ and the two normalised branches as
\begin{align}
\ket{\psi_0}&=\sqrt{1-t^2}\,\sum_n t^{n}\ket{n}_B\ket{n}_{\Bbar},\nonumber\\
\ket{\psi_1}&=(1-t^2)\sum_n\sqrt{n+1}\,t^{n}\ket{n{+}1}_B\ket{n}_{\Bbar},
\label{eq:branches}
\end{align}
where $t=\tanh r$. Tracing out $\Bbar$ from
$\ketbra{B}{B}$ produces four operator-valued blocks on $\mathcal{H}_B$, one for
each pair of Alice indices
\begin{equation}\label{rhoAB_bob}
\rho_{AB}=\sum_{x,x'\in\{0,1\}} c_x c_{x'}\,
\ketbra{x}{x'}_A\otimes\tr_{\Bbar}\ketbra{\psi_x}{\psi_{x'}},
\end{equation}
with $(c_0,c_1)=(\alpha,\beta)$. The three distinct traces are 
\begin{align}
\tr_{\Bbar}\ketbra{\psi_0}{\psi_0}
&=(1-t^2)\sum_n t^{2n}\ketbra{n}{n}_B\;\equiv\;\tau_B,\nonumber\\
\tr_{\Bbar}\ketbra{\psi_1}{\psi_1}
&=(1-t^2)^2\sum_n (n{+}1)\,t^{2n}\ketbra{n{+}1}{n{+}1}_B\;\equiv\;\sigma_B,\nonumber\\
\tr_{\Bbar}\ketbra{\psi_0}{\psi_1}
&=(1-t^2)^{3/2}\sum_n \sqrt{n{+}1}\,t^{2n}\ketbra{n}{n{+}1}_B\;\equiv\;\kappa_B.
\label{eq:threeops}
\end{align}
Their interpretation are as follows: $\tau_B$ is the standard thermal (Bose--Einstein) marginal
of the two-mode squeezed vacuum, with $\tr\tau_B=1$ and mean occupation
$\nn=\sinh^2 r$; $\sigma_B$ is its single-excitation counterpart, also has a unit
trace; and $\kappa_B$ is a \emph{coherence} operator---not a density
matrix---which records the interference between the two branches
\eqref{eq:branches} that survives the trace. It survives precisely because the
two branches carry the same $\Bbar$ occupation ($n$ in both), so the environment
$\Bbar$ does not fully distinguish them; this partial distinguishability is
quantified by $\kappa_B$ and will control every off-diagonal element below. The
$A{:}\Bbar$ reductions follow from the same three traces with the roles of $B$
and $\Bbar$ exchanged, giving 
$\tilde\sigma_B,\tilde\kappa_B$ described in Appendix~\ref{app:rho}. In the
fermionic problem \cite{Basak2023} the analogues of
\eqref{eq:threeops} truncate at $n=0$ by Pauli exclusion; the entire technical
novelty of the bosonic case is that they do not.

\section{Reflected entropy and canonical purification}
\label{sec:reflected}
Any density matrix is Hermitian and positive semidefinite, hence diagonalizable
\begin{equation}
\rho_{XY}=\sum_k p_k\,\ketbra{v_k}{v_k}\;,\qquad p_k\ge0,\quad\sum_k p_k=1\;,
\label{eq:eigrho}
\end{equation}
with orthonormal eigenvectors $\ket{v_k}\in\mathcal H_X\otimes\mathcal H_Y$. The
mixedness is precisely the classical uncertainty over which $\ket{v_k}$ one holds.
To purify it, adjoin a mirror system $X^\star Y^\star$ (isomorphic copies of $X$
and $Y$) and correlate the eigenvalue label into it. The \emph{canonical} choice
records the label in the conjugated eigenvectors themselves
\begin{equation}
\ket{\sqrt{\rho_{XY}}}=\sum_k\sqrt{p_k}\,\ket{v_k}_{XY}\,\ket{v_k^\star}_{X^\star Y^\star}\;,
\label{eq:canpur}
\end{equation}
so that the doubled state is manifestly symmetric between the system and its
mirror. Tracing out the mirror copies returns the
original state, because the $\ket{v_k^\star}$ are orthonormal and the cross terms
vanish,
\begin{equation}
\tr_{X^\star Y^\star}\ketbra{\sqrt{\rho_{XY}}}{\sqrt{\rho_{XY}}}
=\sum_k p_k\,\ketbra{v_k}{v_k}=\rho_{XY}\;.
\end{equation}
The amplitudes of \eqref{eq:canpur} in a product
basis are exactly the matrix elements of the operator
\begin{equation}
M\equiv\sqrt{\rho_{XY}}=\sum_k\sqrt{p_k}\,\ketbra{v_k}{v_k}\;.
\end{equation}
Indeed, expanding \eqref{eq:canpur} in the orthonormal product bases
$\ket{x}_X\ket{y}_Y$ and $\ket{x'}_{X^\star}\ket{y'}_{Y^\star}$, we obtain
\begin{equation}
\ket{\sqrt{\rho_{XY}}}
=\sum_{x,y,x',y'} M_{(x,y),(x',y')}\,
\ket{x}_X\ket{y}_Y\ket{x'}_{X^\star}\ket{y'}_{Y^\star}\;,
\label{eq:purstate}
\end{equation}
with $M_{(x,y),(x',y')}=\bra{x}\bra{y}\,M\,\ket{x'}\ket{y'}$. 

The purified state \eqref{eq:purstate} is pure on the four parties
$X,Y,X^\star,Y^\star$. The reflected entropy is the entanglement of the pair
$\{X,X^\star\}$ against the rest $\{Y,Y^\star\}$, which is defined as
\begin{equation}
\begin{aligned}
\SR(X{:}Y)&=-\tr\!\big(\rho_{XX^\star}\log\rho_{XX^\star}\big)\;,\\
\rho_{XX^\star}&=\tr_{YY^\star}\ketbra{\sqrt{\rho_{XY}}}{\sqrt{\rho_{XY}}}\;,
\end{aligned}
\label{eq:SRdef}
\end{equation}
where $\rho_{X X^{\star}}$ is obtained by tracing out $Y Y^{\star}$.\footnote{Throughout
this paper all logarithms are taken to base two, $\log\equiv\log_2$, so that every
entropy, reflected entropy and Markov gap computed below is measured in bits.}

The intuition for why \eqref{eq:SRdef} captures correlation is that if $X$ and
$Y$ were uncorrelated the purification would factorise, leaving $XX^\star$
unentangled; the more $X$ and $Y$ are correlated, the more the mirror
construction reflects that correlation into entanglement between $X$ and its
copy. Two general properties will be used repeatedly. First, the reflected
entropy is sandwiched between the mutual information and twice the smaller
marginal entropy as
\begin{equation}
\min\{2S_X,2S_Y\}\;\ge\;\SR(X{:}Y)\;\ge\;I(X{:}Y)\;,
\label{eq:bounds}
\end{equation}
so that it is sensitive to classical as well as quantum correlations and
vanishes only on product states---and is bounded whenever the marginals are. Second, the amount by which
the lower bound fails to be saturated is itself informative; it defines the
Markov gap analysed in Sec.~\ref{sec:markov}.

In matrix notation the partial trace in \eqref{eq:SRdef} reads
\begin{equation}\label{eq:RhoAAs}
\left(\rho_{X X^{\star}}\right)_{\left(x, x^{\prime}\right),\left(p, p^{\prime}\right)}=\sum_{y, y^{\prime}} M_{(x, y),\left(x^{\prime}, y^{\prime}\right)} M_{(p, y),\left(p^{\prime}, y^{\prime}\right)}^*\;.
\end{equation}
It is natural to read the $X$ indices as labelling \emph{blocks} and the $Y$
indices as entries within a block: define the block $M^{xx'}$ as the operator on
$\mathcal H_Y$ with matrix elements $(M^{xx'})_{yy'}\equiv M_{(x,y),(x',y')}$. Since
$M$ is Hermitian, the blocks obey $M^{x'x}=(M^{xx'})^{\dagger}$ (and each $M^{xx}$
is Hermitian). The double sum in \eqref{eq:RhoAAs} is then, term by term, a
Hilbert--Schmidt (HS) inner product of two blocks
\begin{equation}
(\rho_{XX^\star})_{(x,x'),(p,p')}
=\tr\!\big[(M^{pp'})^{\dagger}M^{xx'}\big]
\equiv\big\langle M^{pp'},M^{xx'}\big\rangle_{\mathrm{HS}}\;,
\label{eq:gram}
\end{equation}
where $\langle A,B\rangle_{\mathrm{HS}}=\tr[A^{\dagger}B]=\sum_{y,y'}A^{*}_{yy'}B_{yy'}$. Thus $\rho_{XX^\star}$ is the \emph{Gram matrix} of the set of blocks
$\{M^{xx'}\}$: its $\big((x,x'),(p,p')\big)$ entry is the overlap of $M^{xx'}$ with
$M^{pp'}$. So, $\rho_{XX^\star}$ is positive semidefinite and its size is set by the
number of blocks, not by $\dim\mathcal H_Y$.

For any Alice-Bob pairs, the first party is a qubit, $X=A$ with $x\in\{0,1\}$, while the
second is Bob's Fock space, $Y=B$. There are then only four blocks
$\{M^{00},M^{01},M^{10},M^{11}\}$, and $\rho_{AA^\star}$ is a $4\times4$ Gram
matrix---finite \emph{whatever} the dimension of $\mathcal H_B$, since Bob's
infinite-dimensional space is entirely absorbed into the traces \eqref{eq:gram}.
This is exactly what renders the bosonic problem tractable.

\subsection{ The GHZ state}
\label{sec:ghz}
 
We begin our analysis with the GHZ state, as it admits an exact, closed-form expression for the reflected entropy. 
Tracing out $\bar B$ and $C$, we obtain the reduced state of $AB$ in a diagonal form as
\begin{equation}
\rho^{(\mathrm{GHZ})}_{AB}=
\begin{pmatrix}\alpha^2\tau_B & 0\\ 0 & \beta^2\sigma_B\end{pmatrix},
\qquad \beta=\sqrt{1-\alpha^2}.
\label{eq:rhoABghz}
\end{equation}
Its square root is therefore diagonal too,
$M=\mathrm{diag}\big(\alpha\sqrt{\tau_B},\,\beta\sqrt{\sigma_B}\big)$, with
vanishing ladder block $M^{01}=0$. The density matrix $\rho_{AA^\star}$ can be computed using \eqref{eq:gram}, which in the basis $\{|00\rangle,|11\rangle,|01\rangle,|10\rangle\}$ reads
\begin{equation}
\begin{aligned}
\rho_{AA^\star}&=\begin{pmatrix}\alpha^2 & \alpha\beta\,g(r)&0&0\\
\alpha\beta\,g(r) & \beta^2 &0 &0\\
0 &0 &0 &0\\
0 &0 &0 &0
\end{pmatrix}\\[2pt]
&=\begin{pmatrix}\alpha^2 & \alpha\beta\,g(r)\\
\alpha\beta\,g(r) & \beta^2
\end{pmatrix}\oplus\begin{pmatrix}0 & 0\\
0 & 0
\end{pmatrix},
\end{aligned}
\label{eq:rhoAAghz}
\end{equation}
with $g(r)=\tr\!\big[\sqrt{\tau_B}\sqrt{\sigma_B}\big]$,
where $\|M^{00}\|^2=\alpha^2\tr\tau_B=\alpha^2$,
$\|M^{11}\|^2=\beta^2\tr\sigma_B=\beta^2$ (both traces are unity), and
$c=\tr[(\beta\sqrt{\sigma_B})(\alpha\sqrt{\tau_B})]=\alpha\beta\,g$. Since
$\tau_B,\sigma_B$ are diagonal in the Fock basis \eqref{eq:threeops}, the term $g(r)$ becomes
\begin{equation}
g(r)=\tr\!\big[\sqrt{\tau_B}\sqrt{\sigma_B}\big]
=(1-t^2)^{3/2}\sum_{m\ge1}\sqrt{m}\,t^{\,2m-1}\;,
\label{eq:g}
\end{equation}
with $t\equiv\tanh r$.
Physically $g$ is an \emph{affinity}: it quantifies how distinguishable the two
Bob states correlated with $\ket0_A$ (thermal $\tau_B$) and $\ket1_A$
(single-excitation $\sigma_B$) are, with $g\le1$ and $g=1$ only if they coincide.
The eigenvalues of \eqref{eq:rhoAAghz} are
\begin{equation}
    \begin{aligned}
    \lambda_\pm&=\tfrac12\Big[1\pm\sqrt{(\alpha^2-\beta^2)^2+4\alpha^2\beta^2g^2}\Big]\;,\\
    \lambda_3&=\lambda_4=0\;.
\end{aligned}
\end{equation}
For the case of the maximally entangled state ($\alpha=1/\sqrt2$), they reduce to $\lambda_\pm=(1\pm g)/2$, which yields the reflected entropy as
\begin{equation}
\SR^{(\mathrm{GHZ})}(A{:}B)=-\tfrac{1+g}{2}\log\tfrac{1+g}{2}
-\tfrac{1-g}{2}\log\tfrac{1-g}{2}.
\label{eq:ghzSR}
\end{equation}
The infinite-acceleration limit follows analytically. As $r\to\infty$, set
$t^2=e^{-\epsilon}$ with $\epsilon\to0^+$; the sum in \eqref{eq:g} is dominated by
large $m$, $\sum_m\sqrt{m}\,e^{-\epsilon m}\to\Gamma(\tfrac32)\epsilon^{-3/2}
=\tfrac{\sqrt\pi}{2}\epsilon^{-3/2}$, while $(1-t^2)^{3/2}\simeq\epsilon^{3/2}$,
so the $\epsilon$-dependence cancels and we obtain
\begin{equation}
\begin{aligned}
g&\xrightarrow{r\to \infty}\frac{\sqrt\pi}{2},\\
\SR^{(\mathrm{GHZ})}(A{:}B)&\xrightarrow{r\to \infty} H_2\!\Big(\tfrac{1+\sqrt\pi/2}{2}\Big)=0.315 ,
\end{aligned}\;
\label{eq:ghzlimit}
\end{equation}
with $H_2(x)=-x\log x-(1-x)\log(1-x)$ the binary entropy. We notice that $\SR^{(\mathrm{GHZ})}(A{:}B)$ does not vanish at
infinite acceleration; it decreases monotonically from $1$ to a finite
saturation value, because the affinity $g$ saturates at $\sqrt\pi/2<1$.

The reflected entropy for the $A{:}\Bbar$ pair follows from the analogous construction, and for the
GHZ state it can be given in closed form as well. Tracing out $B$ and $C$, we obtain
\begin{equation}
\begin{aligned}
\rho^{(\mathrm{GHZ})}_{A\Bbar}&=
\begin{pmatrix}\alpha^2\tau_B & 0\\ 0 & \beta^2\tilde\sigma_B\end{pmatrix},\\
\tilde\sigma_B&=(1-t^2)^2\sum_{n\ge0}(n{+}1)\,t^{2n}\ketbra{n}{n},
\end{aligned}
\label{eq:rhoABbarghz}
\end{equation}
where $\tilde\sigma_B$ is the anti-Bob partner of $\sigma_B$. The reflected entropy is
therefore given by \eqref{eq:ghzSR} with $g$ replaced by the partner affinity
\begin{equation}
\tilde g(r)=\tr\!\big[\sqrt{\tau_B}\sqrt{\tilde\sigma_B}\big]
=(1-t^2)^{3/2}\sum_{n\ge0}\sqrt{n{+}1}\,t^{\,2n}\;,
\label{eq:gtilde}
\end{equation}
and comparing \eqref{eq:g} with \eqref{eq:gtilde} term by term (set $m=n+1$)
gives the exact relation
\begin{equation}
g(r)=\tanh r\;\tilde g(r)\;,
\label{eq:ggtilde}
\end{equation}
so that a single function determines both pairings. Two consequences follow
without further computation. In the inertial limit $t\to0$ one has
$g\simeq t$, hence $\tilde g\to1$ and
$\SR^{(\mathrm{GHZ})}(A{:}\Bbar)\to H_2(1)=0$, as it must: the second wedge
decouples and carries no correlation. At infinite acceleration $t\to1$, so
\eqref{eq:ggtilde} forces $\tilde g\to g\to\sqrt\pi/2$ and
\begin{equation}
\SR^{(\mathrm{GHZ})}(A{:}\Bbar)\;\longrightarrow\;
H_2\!\Big(\tfrac{1+\sqrt\pi/2}{2}\Big)=0.315\ ,
\label{eq:ghzABbarlimit}
\end{equation}
the \emph{same} saturation value as $\SR^{(\mathrm{GHZ})}(A{:}B)$ in \eqref{eq:ghzlimit}.
The eq~\eqref{eq:ggtilde} thus provides an analytic result, for the GHZ state, of
the convergence of the two Alice pairings observed numerically for all three
states in Sec.~\ref{sec:results}: the factor $\tanh r$ that distinguishes them is
precisely the quantity that tends to unity as the acceleration diverges, which is
the statement that $B$ and $\Bbar$ become statistically indistinguishable. The
two curves approach the common floor from opposite sides---$\SR(A{:}B)$
decreasing from one, $\SR(A{:}\Bbar)$ increasing from zero---as depicted in
Fig.~\ref{fig:saturation}(c).

\subsection{The inter-wedge pair \texorpdfstring{$B{:}\Bbar$}{B:Bbar}}
\label{sec:bbbar}

The construction below is not specific to the GHZ state: it applies to all three
states, which differ only in the weights $p_0,p_1$ appearing in
\eqref{eq:rhoBBstar}.

For the reflected entropy of $B$ and $\Bbar$, both parties live in an infinite-dimensional Fock space, so the qubit reduction fails. Tracing
out $A$ (and $C$) leaves a \emph{rank-two} mixture of two orthonormal branches, we obtain
$\rho_{B\Bbar}=p_0\ketbra{\psi_0}{\psi_0}+p_1\ketbra{\psi_1}{\psi_1}$ (see appendix \ref{app:rho}). Carrying out the canonical purification yields the exact operator identity
\begin{equation}
\rho_{BB^\star}=p_0\,\tau_B\!\otimes\!\tau_B+p_1\,\sigma_B\!\otimes\!\sigma_B
+\sqrt{p_0p_1}\big(\kappa_B\!\otimes\!\kappa_B+\kappa_B^{\dagger}\!\otimes\!\kappa_B^{\dagger}\big).
\label{eq:rhoBBstar}
\end{equation}
Every term preserves $d\equiv n_B-n_{B^\star}$, so $\rho_{BB^\star}$ is
block-diagonal in $d$ and tridiagonal within each sector; diagonalising the
sectors independently makes the spectrum accessible at the large cutoffs
 demanded by large acceleration. Since the marginal entropy
grows as $S(B)\sim2r\log_2e$ and both sides of \eqref{eq:bounds} inherit this growth, which reads
\begin{equation}
\SR(B{:}\Bbar)\sim4r\log_2e\qquad(r\to\infty),
\end{equation}
in stark contrast to the fermionic case, where $B,\Bbar$ are qubits, and
$\SR(B{:}\Bbar)$ saturates to a finite value at large $r$.

\subsection{ The Bell state}
\label{sec:bellworked}

In this section we derive the reflected entropy of the Bell state in a
semi-analytic closed form. Unlike the GHZ case (Sec.~\ref{sec:ghz}), the Bell
state retains genuine Alice--Bob coherence after the trace over $\Bbar$: the
off-diagonal block $\kappa_B$ survives, the three operators
$\tau_B,\sigma_B,\kappa_B$ do not commute, and $\sqrt{\rho_{AB}}$ can no longer
be taken block by block in Alice's basis. The strategy of this section is to
trade Alice's basis for a conserved charge. We proceed in five steps:
(i) identify a $U(1)$ charge conserved by $\rho_{AB}$, which splits the
infinite-dimensional operator into an infinite family of $2\times2$ blocks;
(ii) square-root each block with the elementary $2\times2$ formula;
(iii) reassemble the results into the four Fock-space blocks
$\{M^{00},M^{01},M^{10},M^{11}\}$ of $M=\sqrt{\rho_{AB}}$;
(iv) evaluate the Gram matrix \eqref{eq:gram} from the sector data and
diagonalize the resulting $4\times4$; and
(v) observe that each sector is in fact \emph{rank one}, which collapses three
of the four Gram entries to exact closed forms. A final subsection extends the
construction to the W state, where the sectors are rank two and the
computation remains sector-by-sector numerical.

Tracing $\Bbar$ out of the accelerated Bell state \eqref{eq:bellsub} leaves
$\rho_{AB}$ built from the three operators $\tau_B,\sigma_B,\kappa_B$ derived in
eq.~\eqref{eq:threeops}, which in the basis $\{\ket{0},\ket{1}\}_A$ reads
\begin{equation}
\rho_{AB}=
\begin{pmatrix}\alpha^2\tau_B & \alpha\beta\,\kappa_B\\
\alpha\beta\,\kappa_B^\dagger & \beta^2\sigma_B\end{pmatrix},
\qquad \beta\equiv\sqrt{1-\alpha^2}\;.
\label{eq:rhoABbellmain}
\end{equation}
The diagonal blocks are the branch marginals weighted by Alice's probabilities, and the
off-diagonal block is Alice's coherence dressed by the branch overlap
$\kappa_B$.

Directly square-rooting \eqref{eq:rhoABbellmain} looks intractable, since
$\tau_B,\sigma_B,\kappa_B$ do not commute. The obstacle is removed by a
conserved charge. Consider the Hermitian operator
\begin{equation}
\hat q\equiv \mathbb I_A \otimes \hat n_B-\hat n_A \otimes\mathbb I_B ,
\label{eq:qdef}
\end{equation}
with $\hat n_A=\ketbra{1}{1}_A$ and $\hat n_B=\sum_n n\ketbra{n}{n}_B$ the
number operators; every product state is an eigenstate,
$\hat q\,\ket{x}_A\ket{n}_B=(n-x)\ket{x}_A\ket{n}_B$, and we label it by the
integer eigenvalue $q=n-x$ (the shorthand ``$q=n_B-n_A$''). The claim is that
$[\rho_{AB},\hat q]=0$; equivalently, since
$\bra{\phi_a}[\rho_{AB},\hat q]\ket{\phi_b}=(q_b-q_a)\bra{\phi_a}\rho_{AB}\ket{\phi_b}$
in a $\hat q$-eigenbasis, $\rho_{AB}$ has no matrix elements between states of
different charge. One checks this term by term in
\eqref{eq:rhoABbellmain}: the diagonal blocks $\alpha^2\tau_B$ and
$\beta^2\sigma_B$ are diagonal in $n_B$ and preserve $q$ trivially, while the
off-diagonal block $\alpha\beta\kappa_B$, sitting in the $\ketbra{0}{1}_A$ slot,
sends $\ket{1}_A\ket{n{+}1}_B\to\ket{0}_A\ket{n}_B$, for which
$q=(n{+}1)-1=n$ before and $q=n-0=n$ after: lowering $n_B$ by one is exactly
compensated by lowering $n_A$ by one. Physically, this charge is the descendant
of Bogoliubov charge conservation---each Alice excitation is created together
with exactly one extra $B$ quantum---after $\Bbar$ has been traced out.

Consequently $\rho_{AB}$ is block diagonal in the charge sectors
$q=0,1,2,\dots$, and---because $A$ is a qubit---each sector contains exactly two
states, $\{\ket{0,q}_{AB},\,\ket{1,q{+}1}_{AB}\}$, so each block is only
$2\times2$. Thus \eqref{eq:rhoABbellmain} can be expressed as  $\rho_{AB}\equiv\bigoplus_{q}\rho^{(q)}$ with
\begin{equation}
\begin{aligned}
\rho^{(q)}&=(1-t^2)t^{2q}
\begin{pmatrix}\alpha^2 & \alpha\beta\sqrt{1-t^2}\sqrt{q{+}1}\\[2pt]
\alpha\beta\sqrt{1-t^2}\sqrt{q{+}1} & \beta^2(1-t^2)(q{+}1)\end{pmatrix}\;,\\
&\equiv\begin{pmatrix}a_q & b_q\\ b_q & d_q\end{pmatrix}.
\end{aligned}
\label{eq:rhoq}
\end{equation}
An infinite operator has become an infinite family of ordinary $2\times2$
matrices, one for each value of $q$.

The square root of a real, symmetric, positive $2\times2$ matrix is simply
\begin{equation}
\begin{aligned}
\sqrt{\begin{pmatrix}a&b\\b&d\end{pmatrix}}
&=\frac{1}{s}\begin{pmatrix}a+\Delta & b\\ b & d+\Delta\end{pmatrix},\\
\Delta&=\sqrt{ad-b^2},\qquad s=\sqrt{a+d+2\Delta}.
\end{aligned}
\label{eq:sqrt2x2}
\end{equation}
Applying \eqref{eq:sqrt2x2} to \eqref{eq:rhoq}, we obtain
\begin{equation}
\sqrt{\rho^{(q)}}=\left(\begin{array}{ll}
m_{11}^{(q)} & m_{12}^{(q)} \\
m_{12}^{(q)} & m_{22}^{(q)}
\end{array}\right), \quad \quad m_{11}^{(q)}=\frac{a_q+\Delta_q}{s_q}, \quad m_{22}^{(q)}=\frac{d_q+\Delta_q}{s_q}, \quad m_{12}^{(q)}=\frac{b_q}{s_q},
\end{equation}
with $\Delta_q=\sqrt{a_q d_q-b_q^2}, s_q=\sqrt{a_q+d_q+2 \Delta_q}$. Thus each $\rho^{(q)}$ of \eqref{eq:rhoq} gives three
numbers per sector, $m^{(q)}_{11},m^{(q)}_{12},m^{(q)}_{22}$, which assemble
into the four Fock-space blocks of $M=\sqrt{\rho_{AB}}$ as
\begin{equation}
\begin{aligned}
(M^{00})_{qq}&=m^{(q)}_{11},\qquad
(M^{11})_{q+1,q+1}=m^{(q)}_{22},\\
(M^{01})_{q,q+1}&=m^{(q)}_{12}.
\end{aligned}
\label{eq:Mblocksassembled}
\end{equation}
Thus $M^{00}$ is diagonal, $M^{11}$ is diagonal but shifted up by one Fock
level, and $M^{01}=(M^{10})^\dagger$ is a  ladder operator; no cross
terms between different $q$'s appear anywhere, because the charge $q$ forbids
them.

In sec~\ref{sec:reflected}, we showed that
$(\rho_{AA^\star})_{(x,x'),(p,p')}=\tr[(M^{pp'})^\dagger M^{xx'}]$ in \eqref{eq:gram}. Using
\eqref{eq:Mblocksassembled}, all four needed overlaps reduce to convergent sums
over sectors as
\begin{align}
\|M^{00}\|^2 &=\operatorname{Tr}\left[\left(M^{00}\right)^{\dagger} M^{00}\right]=\sum_{q\ge0}\big(m^{(q)}_{11}\big)^2, &
\|M^{11}\|^2 &=\operatorname{Tr}\left[\left(M^{11}\right)^{\dagger} M^{11}\right]=\sum_{q\ge0}\big(m^{(q)}_{22}\big)^2, \label{eq:gramsumsA}\\
\|M^{01}\|^2 &=\operatorname{Tr}\left[\left(M^{01}\right)^{\dagger} M^{01}\right]=\sum_{q\ge0}\big(m^{(q)}_{12}\big)^2, &
c &=\operatorname{Tr}\left[\left(M^{11}\right)^{\dagger} M^{00}\right]=\sum_{q\ge0} m^{(q)}_{22}\,m^{(q+1)}_{11}. \label{eq:gramsumsB}
\end{align}
The first three sums are purely \emph{within} a sector; $c=\tr[(M^{11})^\dagger
M^{00}]$ is the one genuinely off-diagonal object, and it couples
\emph{consecutive} sectors $q$ and $q{+}1$, because $M^{00}$ carries weight at
Fock level $n=q$ while $M^{11}$ carries it at $n=q{+}1$. Physically, $c$ is the
residual quantum interference between the two Bogoliubov branches
\eqref{eq:vac}--\eqref{eq:one}.

It remains to see which of the sixteen Gram entries survive. By
\eqref{eq:Mblocksassembled} each block is a ladder operator of definite Fock
shift: $M^{00}$ and $M^{11}$ shift $n_B$ by $0$, $M^{01}$ lowers it by one, and
$M^{10}$ raises it by one. Under a trace only the shift-zero (diagonal) part of
an operator contributes, so $\tr[(M^{pp'})^\dagger M^{xx'}]$ vanishes unless the
two shifts cancel, i.e.\ unless $x-x'=p-p'$. The four diagonal norms trivially
survive; among the off-diagonal entries the only shift-matched pair is
$\{M^{00},M^{11}\}$, producing the single coherence $c$ (all entries pairing a
diagonal block with a ladder block, or $M^{01}$ with $M^{10}$, are traceless).
In the ordered basis $\{\ket{00},\ket{11},\ket{01},\ket{10}\}_{AA^\star}$, $\rho_{AA^\star}$ \eqref{eq:gram} is therefore given as 
\begin{equation}
\rho_{AA^\star}=
\begin{pmatrix}\|M^{00}\|^2 & c\\ c^* & \|M^{11}\|^2\end{pmatrix}
\oplus
\begin{pmatrix}\|M^{01}\|^2 & 0\\ 0 & \|M^{01}\|^2\end{pmatrix},
\label{eq:rhoAAbellmain}
\end{equation}
where we used $\|M^{10}\|^2=\|M^{01}\|^2$. The four diagonal weights sum to
$\tr[M^\dagger M]=\tr\rho_{AB}=1$, and 
the eigenvalues are
\begin{equation}
\begin{aligned}
\lambda_\pm&=\frac{\|M^{00}\|^2+\|M^{11}\|^2}{2}
\pm\sqrt{\Big(\tfrac{\|M^{00}\|^2-\|M^{11}\|^2}{2}\Big)^2+c^2},\\
\lambda_0&=\|M^{01}\|^2 \quad \text{(two folds)},
\end{aligned}
\end{equation}
and the reflected entropy is 
\begin{equation}
\SR^{(B)}(A{:}B)=-\lambda_+\log\lambda_+-\lambda_-\log\lambda_--2\lambda_0\log\lambda_0.
\label{eq:SRbellfinal}
\end{equation}

\subsubsection{Semi-analytic closed form}
\label{sec:bellanalytic}

The sector construction can be pushed further, to a nearly closed-form answer.
The crucial observation is that each block $\rho^{(q)}$ in \eqref{eq:rhoq} is
\emph{rank one}: a short computation gives
\begin{equation}
\begin{aligned}
\det\rho^{(q)}&=a_q d_q-b_q^2\\
&=\alpha^2\beta^2(1-u)^3(q{+}1)u^{2q}\\
&\quad-\alpha^2\beta^2(1-u)^3(q{+}1)u^{2q}=0,
\end{aligned}
\label{eq:rank1}
\end{equation}
with $u\equiv t^2=\tanh^2 r$. Physically, for fixed $q$ both Bogoliubov branches
in \eqref{eq:bellsub} carry the \emph{same} anti-Bob occupation $n_{\Bbar}=q$,
so tracing out $\Bbar$ does not decohere them and the Alice--Bob state remains
pure within each charge sector. For a rank-one positive matrix $P$ one has
$\sqrt{P}=P/\sqrt{\tr P}$, so the sector roots are simply
$m^{(q)}_{ij}=\rho^{(q)}_{ij}/\sqrt{w_q}$ with $w_q=a_q+d_q$.

Assembled into Fock-space operators via \eqref{eq:Mblocksassembled}, and writing
$D_q\equiv\alpha^2+\beta^2(1-t^2)(q{+}1)$ so that $w_q=(1-t^2)t^{2q}D_q$, the four
blocks of $M=\sqrt{\rho_{AB}}$ are given in closed form by
\begin{equation}
\begin{aligned}
M^{00}&=\sqrt{1-t^2}\,\sum_{q\ge0}\frac{\alpha^2\,t^{q}}{\sqrt{D_q}}\,\ketbra{q}{q},\\
M^{11}&=(1-t^2)^{3/2}\sum_{q\ge0}\frac{\beta^2(q{+}1)\,t^{q}}{\sqrt{D_q}}\,\ketbra{q{+}1}{q{+}1},\\
M^{01}&=(1-t^2)\sum_{q\ge0}\frac{\alpha\beta\sqrt{q{+}1}\,t^{q}}{\sqrt{D_q}}\,\ketbra{q}{q{+}1},
\end{aligned}
\label{eq:bellblocks}
\end{equation}
and $M^{10}=(M^{01})^{\dagger}$. These realize explicitly the structure
anticipated in Sec.~\ref{sec:reflected}: $M^{00},M^{11}$ carry the Fock-diagonal
form of $\sqrt{\tau_B},\sqrt{\sigma_B}$ and $M^{01}$ the one-step ladder form of
$\kappa_B$, each dressed by the sector factor $1/\sqrt{D_q}$ that encodes the
non-commutativity of the three operators. (For GHZ the coherence is absent from
the outset, $\kappa_B=0$; the sectors are then uncoupled rather than rank one, and
the blocks reduce to the elementary roots $M^{00}=\alpha\sqrt{\tau_B}$,
$M^{11}=\beta\sqrt{\sigma_B}$, $M^{01}=0$ of Sec.~\ref{sec:ghz}.)

The three diagonal Gram sums \eqref{eq:gramsumsA}--\eqref{eq:gramsumsB} then
collapse to a single \emph{Lerch transcendent}
$\Phi_\ast\equiv\Phi(u,1,1+\nu)=\sum_{q\ge0}u^q/(q{+}1{+}\nu)$, with
$\nu\equiv\alpha^2/[\beta^2(1-u)]$:
\begin{equation}
\begin{aligned}
\|M^{00}\|^2&=\frac{\alpha^4}{\beta^2}\,\Phi_\ast,\qquad
\|M^{01}\|^2=\alpha^2-\|M^{00}\|^2,\\
\|M^{11}\|^2&=\beta^2-\alpha^2+\|M^{00}\|^2.
\end{aligned}
\label{eq:bellclosed}
\end{equation}
The last two relations follow from the elementary sums
$\sum_q u^q=(1-u)^{-1}$ and $\sum_q(q{+}1)u^q=(1-u)^{-2}$; they guarantee the
normalisation $\|M^{00}\|^2+\|M^{11}\|^2+2\|M^{01}\|^2=1$ automatically. Only the
interference coherence resists a closed form, which reads
\begin{equation}
c=\alpha^2(1-u)\sqrt{u}\sum_{q\ge0}
\frac{(q{+}1)\,u^{q}}{\sqrt{(q{+}1{+}\nu)(q{+}2{+}\nu)}},
\label{eq:bellcoherence}
\end{equation}
because the square root of the product of two consecutive shifted factors is not
a standard hypergeometric kernel. Thus $\SR(A{:}B)$ is \emph{semi-analytic}: it
reduces to the $2\times2$ eigenvalue problem \eqref{eq:rhoAAbellmain} in which
three of the four entries are exact closed forms and only the scalar
\eqref{eq:bellcoherence}---a rapidly convergent, bounded sum---need be evaluated
numerically. 

The saturation value of $\SR(A{:}B)$ likewise admits an exact expression. As
$r\to\infty$, $u\to1$ and $\nu\to\infty$; converting the sums
\eqref{eq:bellclosed}--\eqref{eq:bellcoherence} to integrals (with $x=(1-u)q$)
brings in the exponential integral through
\begin{equation}
\mathcal{E}(k)\equiv e^{k}E_1(k)=\int_0^\infty\!\frac{e^{-x}}{x+k}\,dx,
\qquad k\equiv\frac{\alpha^2}{\beta^2}.
\end{equation}
The Gram data acquire the limits
$\|M^{00}\|^2\to\tfrac{\alpha^4}{\beta^2}\mathcal{E}(k)$ and, notably we obtain
\begin{equation}
c\;\to\;\alpha^2-\|M^{00}\|^2=\|M^{01}\|^2 ,
\end{equation}
i.e. the branch-interference coherence merges exactly with the $\lambda_0$ block.
For the maximally entangled Bell state ($\alpha^2=\tfrac12$, $k=1$) the four
eigenvalues of $\rho_{AA^\star}$, with $\mathcal{E}\equiv e\,E_1(1)$, are
\begin{equation}
\Big\{\tfrac12,\ \ \mathcal{E}-\tfrac12,\ \ \tfrac{1-\mathcal{E}}{2},\ \ \tfrac{1-\mathcal{E}}{2}\Big\},
\end{equation}
giving the closed-form saturation floor as
\begin{equation}
\SR^{(B)}(A{:}B)\big|_{r\to\infty}
=-\tfrac12\log\tfrac12-\big(\mathcal{E}-\tfrac12\big)\log\big(\mathcal{E}-\tfrac12\big)-(1-\mathcal{E})\log\tfrac{1-\mathcal{E}}{2}\;,
\label{eq:bellfloor}
\end{equation}
with $\mathcal{E}=e\,E_1(1)$, we obtain $\SR^{(B)}(A{:}B)\big|_{r\to\infty}  =1.7572$. This is the
Bell counterpart of the GHZ floor $H_2(\tfrac12(1+\sqrt\pi/2))$ of \eqref{eq:ghzlimit}, and matches the saturation plotted numerically in Fig.~\ref{fig:saturation}(a). The same construction at general $\alpha$ gives the
saturation value through $\|M^{00}\|^2_\infty=\tfrac{\alpha^4}{\beta^2}\mathcal{E}(k)$ and
$c_\infty=\|M^{01}\|^2_\infty=\alpha^2-\|M^{00}\|^2_\infty$.

The reflected entropy for the remaining two pairs can be computed anlogously. For $A{:}\Bbar$ one traces
out $B$ (rather than $\Bbar$) in \eqref{eq:bellsub}, which replaces $\sigma_B$ by
its unshifted partner $\tilde\sigma_B$ and $\kappa_B$ by $t\,\tilde\kappa_B$
(Appendix~\ref{app:rho}). Since $\tilde\kappa_B$ \emph{raises} the Fock index
where $\kappa_B$ lowers it, the conserved charge changes sign, $\hat q$ being
replaced by $\hat n_{\Bbar}+\hat n_A$, and the sectors become
$\{\ket{0,q},\ket{1,q{-}1}\}$ for $q\ge1$ together with the one-dimensional
sector $\{\ket{0,0}\}$. Their determinants again vanish identically, so the
sectors remain rank one and the closed forms of
Sec.~\ref{sec:bellanalytic} carry over verbatim; only the Fock weights entering the Gram sums change. For
$B{:}\Bbar$ the qubit reduction is unavailable, but the state is the rank-two
mixture treated in Sec.~\ref{sec:bbbar}, so \eqref{eq:rhoBBstar} applies directly
with $p_0=\alpha^2$ and $p_1=1-\alpha^2$, and the spectrum follows from the
tridiagonal blocks labelled by $d=n_B-n_{B^\star}$. We have verified that the
sector construction reproduces the direct numerical purification for all three
pairings; the resulting curves for the reflected entropy are shown in Fig.~\ref{fig:reflected}.

\subsection{The W state}
\label{sec:wstate}
 
The same sector strategy applies to the W state, with two instructive
differences. Tracing out $\Bbar$ and $C$ gives (see eq.~\eqref{eq:rhoABwApp})
\begin{equation}
\rho^{(W)}_{AB}=
\begin{pmatrix}\alpha^2\tau_B+\gamma^2\sigma_B & \alpha\gamma\,\kappa_B^{\dagger}\\
\alpha\gamma\,\kappa_B & \alpha^2\tau_B\end{pmatrix},
\qquad \gamma\equiv\sqrt{1-2\alpha^2}.
\label{eq:rhoABw}
\end{equation}
First, the conserved charge changes sign: the coherence now sits in the
$\ketbra{0}{1}_A$ slot as the \emph{raising} operator $\kappa_B^\dagger$, sending
$\ket{1}_A\ket{n}_B\to\ket{0}_A\ket{n{+}1}_B$, so the conserved combination is
$\hat q_W=\hat n_B+\hat n_A$ rather than $\hat n_B-\hat n_A$ (the W coherence
ties Alice's excitation to Bob's \emph{thermal} branch rather than to his
excited one). The sectors are $\{\ket{0,q}_{AB},\ket{1,q{-}1}_{AB}\}$ for
$q\ge1$, plus the one-dimensional sector $\{\ket{0,0}\}$, and
$\rho^{(W)}_{AB}$ is again an infinite family of $2\times2$ blocks.
 
Second, the sectors are \emph{rank two}. The determinant of the sector block
evaluates to
\begin{equation}
\det\rho^{(q)}_W=\alpha^4(1-u)^2u^{\,2q-1}\neq0 ,
\label{eq:wdet}
\end{equation}
the $\gamma$-dependent contributions cancelling identically. The physical reason
is the third branch of the W state: the Charlie-excited term
$\alpha\ket{0}_A\ket{0}_M\ket{1}_C$ is orthogonal in Charlie's register to the
two coherent branches, so it deposits the \emph{incoherent} weight
$\alpha^2\tau_B$ onto the Alice-$0$ diagonal of \eqref{eq:rhoABw}. Within each
sector this incoherent admixture lifts the purity that made the Bell sectors
rank one---indeed \eqref{eq:wdet} is precisely (that admixture)$\times$(the
$\alpha^2\tau_B$ diagonal of the Alice-$1$ block).
 
Consequently the rank-one shortcut $\sqrt P=P/\sqrt{\tr P}$ is unavailable and no
Lerch-type closed form arises; one instead square-roots each sector with the
general $2\times2$ formula. Reading \eqref{eq:rhoABw} in the sector basis
$\{\ket{0,q}_{AB},\ket{1,q{-}1}_{AB}\}$, the block is
$\rho^{(q)}_W$ is
\begin{equation}
   \rho^{(q)}_W=\begin{pmatrix}a_q & b_q\\ b_q & d_q\end{pmatrix} \;,
\end{equation}
with
\begin{equation}
\begin{aligned}
a_q&=(1{-}u)u^{q-1}\big[\alpha^2 u+\gamma^2(1{-}u)q\big],\\
d_q&=\alpha^2(1{-}u)u^{q-1},\qquad
b_q=\alpha\gamma(1{-}u)^{3/2}\sqrt{q}\,u^{q-1},
\end{aligned}
\label{eq:wsectordata}
\end{equation}
for $q\ge1$ (the $q=0$ sector contributes the single number $a_0=\alpha^2(1-u)$).
Applying \eqref{eq:sqrt2x2} with
$\Delta_q=\sqrt{\det\rho^{(q)}_W}=\alpha^2(1-u)u^{q-1/2}$ and
$s_q=\sqrt{a_q+d_q+2\Delta_q}$ gives the sector roots
$m^{(q)}_{11}=(a_q+\Delta_q)/s_q$, $m^{(q)}_{22}=(d_q+\Delta_q)/s_q$ and
$m^{(q)}_{12}=b_q/s_q$, together with $m^{(0)}_{11}=\alpha\sqrt{1-u}$. Because the
sector states are $\{\ket{0,q},\ket{1,q{-}1}\}$, these assemble as
$(M^{00})_{q,q}=m^{(q)}_{11}$, $(M^{11})_{q-1,q-1}=m^{(q)}_{22}$ and
$(M^{01})_{q,q-1}=m^{(q)}_{12}$; in particular $M^{01}$ now \emph{raises} $n_B$ by
one, the mirror of the Bell case.
 
The selection rule $x-x'=p-p'$ is unchanged, so $\rho_{AA^\star}$ keeps the block
form \eqref{eq:rhoAAbellmain}, now built from the rank-two sector sums
\begin{equation}
\begin{aligned}
\|M^{00}\|^2&=\sum_{q\ge0}\big(m^{(q)}_{11}\big)^2,\qquad
\|M^{11}\|^2=\sum_{q\ge1}\big(m^{(q)}_{22}\big)^2,\\
\|M^{01}\|^2&=\sum_{q\ge1}\big(m^{(q)}_{12}\big)^2,\qquad
c=\sum_{q\ge0} m^{(q)}_{11}\,m^{(q+1)}_{22}.
\end{aligned}
\label{eq:wgramsums}
\end{equation}
As for Bell, the coherence couples consecutive sectors, but the offset now falls
on $m_{22}$ (the mirrored level dictionary), and only the $\|M^{00}\|^2$ sum
carries the $q=0$ term. The computation is exact sector-by-sector and numerically
trivial: for $\alpha=1/\sqrt3$,  at large
acceleration $S^{(W)}_R({A}:{B})$ saturates at $\approx0.75$, which we depict in Fig.~\ref{fig:saturation}(b).

The other two pairings for the W state follow the same route, with two
differences worth mentioning. For $A{:}\Bbar$, tracing out $B$ and $C$ gives
\begin{equation}
\rho^{(W)}_{A\Bbar}=
\begin{pmatrix}\alpha^2\tau_B+\gamma^2\tilde\sigma_B & \alpha\gamma\,\tilde\kappa_B^{\dagger}\\
\alpha\gamma\,\tilde\kappa_B & \alpha^2\tau_B\end{pmatrix}\;.
\label{eq:rhoABbarw}
\end{equation}
 Since $\tilde\kappa_B$ raises
the Fock index, the conserved charge is now $\hat n_{\Bbar}-\hat n_A$, and the
sectors are $\{\ket{0,q},\ket{1,q{+}1}\}$ for $q\ge0$ together with the
one-dimensional sector $\{\ket{1,0}\}$---the mirror image of the $A{:}B$
decomposition, in which the one-dimensional sector sits at $\ket{0,0}$. Thus for $A{:}B$ pair, these
sectors are rank two, so the general $2\times2$ root is again required and no
closed form results. For $B{:}\Bbar$, the three branches of \eqref{eq:wsub} are
mutually orthogonal in the $A$ and $C$ registers, so tracing them out leaves the
rank-two mixture $\rho_{B\Bbar}=2\alpha^2\ketbra{\psi_0}{\psi_0}
+(1-2\alpha^2)\ketbra{\psi_1}{\psi_1}$ as discussed in Sec~\ref{sec:bbbar}: the two branches carrying $\ket{\psi_0}$
add incoherently and their weights combine. Eq.~\eqref{eq:rhoBBstar}
therefore applies with $p_0=2\alpha^2$ and $p_1=1-2\alpha^2$. We have checked
that the sector construction reproduces the direct purification for both
pairings; the resulting curves for the reflected entropy are shown in Fig.~\ref{fig:reflected}.

\subsection{Reflected entropy results}
\label{sec:results}

\begin{figure*}[t]
\centering
\includegraphics[width=\textwidth]{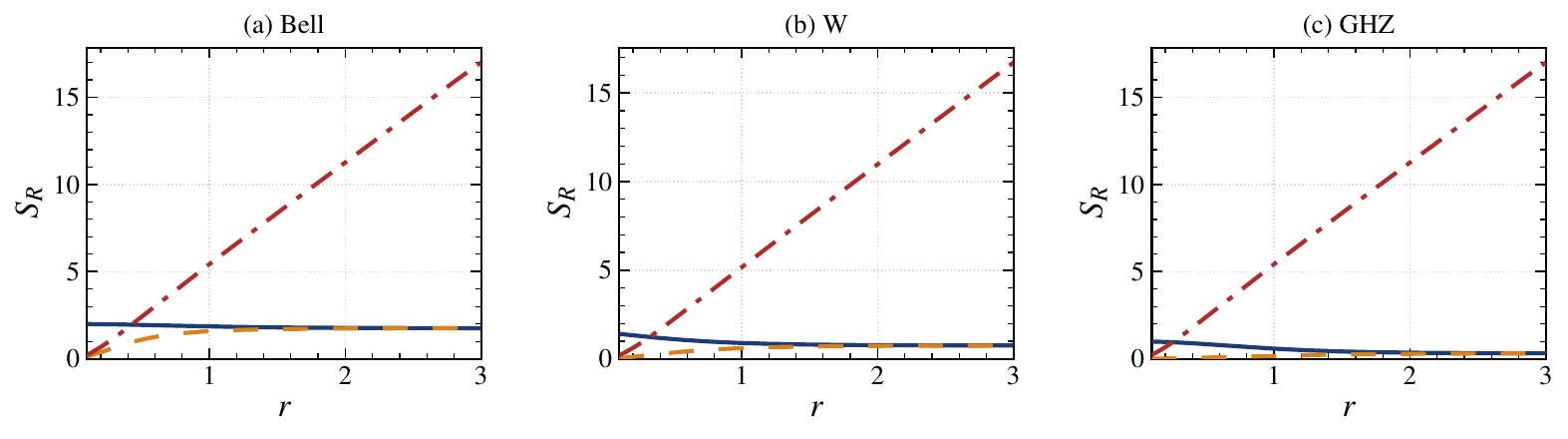}
\caption{Reflected entropy of the three pairings as a function of the squeezing
parameter $r$, for (a) the Bell, (b) the W and (c) the GHZ state, at
$\alpha=1/\sqrt2$ for Bell and GHZ and $\alpha=1/\sqrt3$ for W.  The three curves are: $S_R(A:B)$ (solid blue), $S_R(A:\bar B)$ (orange dashed) and $S_R(B:\bar B)$  (red dot-dashed).}
\label{fig:reflected}
\end{figure*}

In Fig.~\ref{fig:reflected}, we depict the three reflected entropies as a function $r$ at the
maximally entangled values ($\alpha=1/\sqrt2$ for Bell/GHZ, $1/\sqrt3$ for W).
\begin{figure*}[t]
\centering
\includegraphics[width=\textwidth]{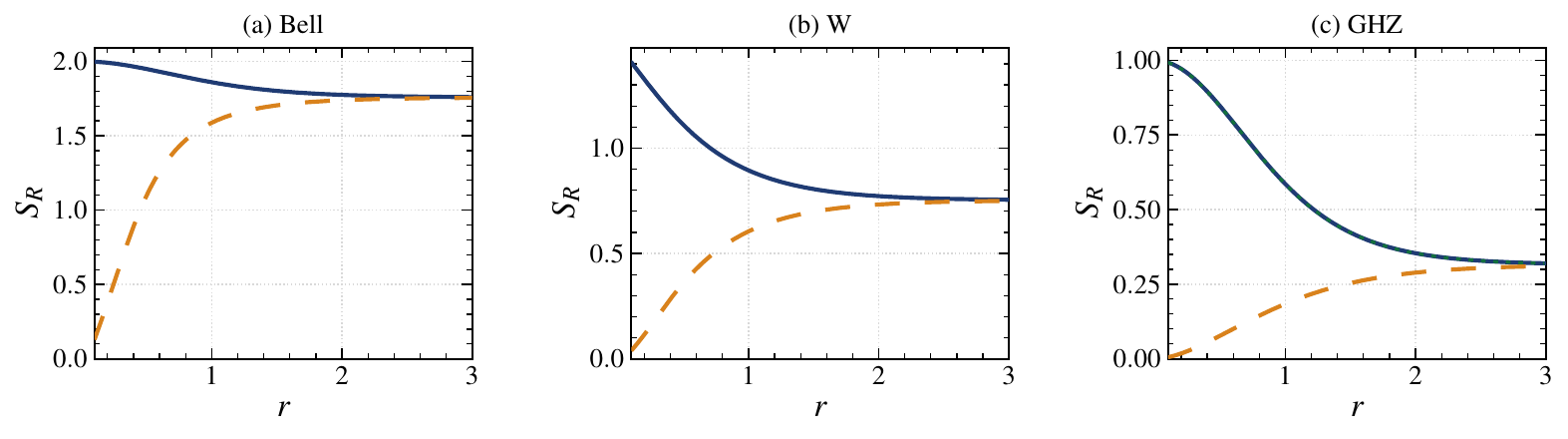}
\caption{Saturation of the reflected entropy for the two Alice pairings a function of the squeezing
parameter $r$, for (a) Bell, (b) W and (c) GHZ, with the same parameters as
Fig.~\ref{fig:reflected}. The two curves are: $S_R(A:B)$ (solid blue) and  $S_R(A:\bar B)$ (orange dashed).}
\label{fig:saturation}
\end{figure*}
We observe that the Alice--Bob reflected entropy degrades with acceleration---most steeply for
GHZ, most gently for Bell---but it does \emph{not} decay to zero. It saturates at
a finite floor: $1.757$ (Bell), $0.752$ (W), and $0.315$
(GHZ, eq.~\eqref{eq:ghzlimit}); see Fig.~\ref{fig:saturation}. 
The boundedness is guaranteed a priori by \eqref{eq:bounds}, since
$\SR(A{:}B)\le2S_A\le2$ with $A$ a qubit; the nonzero floor reflects that
reflected entropy, unlike an entanglement measure, does not vanish on
classically correlated states: it is bounded below by the mutual information and
vanishes only if $\rho_{AB}$ is a product state. Although bosonic
entanglement is degraded away at large $r$, the classical correlation between
Alice and Bob survives, with residual distinguishability $g_\infty=\sqrt\pi/2<1$. Two known
results support this reading. First, for scalar fields the classical correlation
is essentially insensitive to acceleration \cite{MartinMartinezLeon2010}, so any
{quantity bounded below by the mutual information is expected to inherit a
saturation floor from it. Second, the
survival of \emph{quantum} correlation beyond entanglement in this same bosonic
setting was established for quantum discord by Datta \cite{Datta2009}, who found
a nonzero discord residue at infinite acceleration precisely where distillable
entanglement vanishes. Our saturation values are the reflected-entropy counterparts of that
phenomenon, with the added advantage of exact closed forms
(Secs.~\ref{sec:ghz}, \ref{sec:bellanalytic}).

As $r\to\infty$, $\SR(A{:}B)$ and $\SR(A{:}\Bbar)$ converge to a \emph{common}
value (Fig.~\ref{fig:saturation}): the correlation Alice shares with the
accessible wedge $B$ and with the inaccessible wedge $\Bbar$ become equal once the
two are maximally mixed by the Unruh effect. This mirrors the fermionic statement
that $B$ and $\Bbar$ become indistinguishable at infinite acceleration.

The only unbounded quantity is $\SR(B{:}\Bbar)$, growing as $4r$
as shown in Fig.~\ref{fig:reflected}. This is
the cleanest fermion/boson discriminator: a finite ceiling at unit value for fermions
versus linear divergence for bosons, tied directly to the unbounded thermal
occupation $\nn=\sinh^2 r$ of the Rindler modes. We notice that
$\rho_{B\Bbar}$, hence $\SR(B{:}\Bbar)$, is \emph{identical} for the Bell and GHZ
states at equal $\alpha$, since tracing out $A$ leaves the same rank-two mixture.

\section{Markov gap}
\label{sec:markov}

 As we mentioned earlier, that the reflected entropy is bounded below by the mutual information
eq.~\eqref{eq:bounds}. The amount by which that bound fails to be saturated
defines the \emph{Markov gap} \cite{HaydenParrikarSorce2021}, which is
\begin{equation}
h(X{:}Y)=\SR(X{:}Y)-I(X{:}Y)\;\ge\;0\;.
\label{eq:markovdef}
\end{equation}
Its meaning is sharpest in the
canonically purified state \eqref{eq:purstate}, which is pure on the four parties
$X,Y,X^\star,Y^\star$. There one has the identity
\begin{equation}
h(X{:}Y)=\SR(X{:}Y)-I(X{:}Y)=I(X{:}Y^\star|Y),
\label{eq:markovcmi}
\end{equation}
with $I(X{:}Y|Z)=S_{XZ}+S_{YZ}-S_{Z}-S_{XYZ}$ the conditional mutual information;
\eqref{eq:markovcmi} follows from $S_{XX^\star}=\SR$, the purity of the doubled
state and the mirror symmetry $S_X=S_{X^\star}$, $S_{XY}=S_{X^\star Y^\star}$. Two
consequences fix the interpretation. First, $h\ge0$ is precisely strong subadditivity, so the gap is nonnegative for
any state, and $h=0$ characterises a \emph{quantum Markov chain}
$X\to Y\to Y^\star$: the case in which $X$ is correlated with $Y^\star$ only
through $Y$. Second, by the Fawzi--Renner theorem \cite{FawziRenner2015} a small
conditional mutual information implies the existence of a recovery channel
$\mathcal R_{Y\to YY^\star}$ acting on $Y$ alone that approximately reconstructs
the global state
\begin{equation}
F\big(\rho_{XYY^\star},\,\mathcal R_{Y\to YY^\star}(\rho_{XY})\big)\;\gtrsim\;2^{-h(X{:}Y)} ,
\label{eq:recovery}
\end{equation}
where $F$ represents the fidelity. A large gap therefore means the correlation between Alice
and Bob is \emph{irreducibly tripartite}: no operation performed by Bob alone can
restore what was lost when the third party was traced out. This is the sense in
which $h$ probes structure that the mutual information---a strictly bipartite
quantity---cannot see, and it is why $h$ and $\SR$ carry complementary
information. In the holographic context the same quantity is bounded below by the
number of boundaries of the entanglement-wedge cross section
\cite{HaydenParrikarSorce2021}, consistent with the observation that the
cross section itself requires genuinely tripartite entanglement
\cite{AkersRath2020}; here the relevant tripartition is
$A,B,\Bbar$, so a nonzero gap is a statement about how the Unruh effect
distributes correlation across the horizon.

\subsection{Markov gap Results}
\label{sec:Margap}
 
In Fig.~\ref{fig:markov}, we depict the Markov gap as a function $r$. All of them
\emph{saturate}: like the Alice-pair reflected entropies, and unlike
$\SR(B{:}\Bbar)$, the Markov gaps remain bounded as $r\to\infty$. This is
expected from \eqref{eq:markovcmi}, since the divergent pieces of $\SR$ and $I$
cancel in the difference---the gap is sensitive to the \emph{structure} of the
correlation, not to its magnitude.
 
\begin{figure*}[t]
\centering
\includegraphics[width=\textwidth]{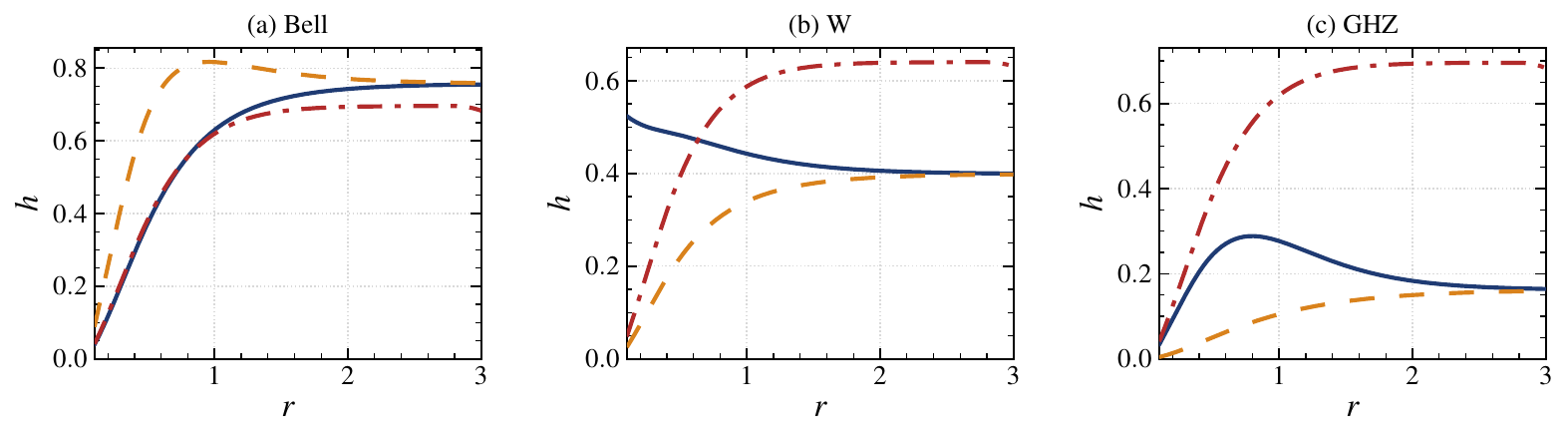}
\caption{The Markov gap $h$ for the three pairings as a function of $r$, for
(a) Bell, (b) W and (c) GHZ, with the same parameters as Fig.~\ref{fig:reflected}. The three curves are:
$h(A{:}B)$ (blue solid), $h(A{:}\Bbar)$  (orange dashed), and $h(B{:}\Bbar)$ (red dot-dashed).}
\label{fig:markov}
\end{figure*}
 
For Bell and GHZ the gaps \emph{rise from zero}. This has a clean origin: at
$r=0$ the state factorises as
$\ket{\Phi}_{AB}\otimes\ket{0}_{\Bbar}$ (Sec.~\ref{sec:setup}), so $\rho_{AB}$ is
pure for Bell and separable for GHZ; in either case the Alice--Bob correlation is
purely bipartite, the Markov condition holds, and $h=0$. Switching on the
acceleration populates $\Bbar$ and \emph{creates} the third party, so any nonzero
gap at $r>0$ is tripartite correlation generated by the Unruh effect alone. Its
persistence at large $r$---$h(A{:}B)\to0.75$ for Bell and $0.16$ for
GHZ---signals genuine three-party structure among $A,B,\Bbar$, paralleling the
fermionic finding that the Markov gap detects correlations invisible to the
residual tangle \cite{Basak2023,Coffman2000}.

The W state stands apart, and instructively so. Unlike Bell and GHZ, it already carries genuinely tripartite correlation in the inertial limit, where $h(A{:}B)\simeq0.57$. Acceleration does not add to this: $h(A{:}B)$
\emph{decreases} monotonically, settling at $\simeq0.40$, while the two channels
involving the causally disconnected wedge grow from zero---$h(A{:}\Bbar)$
to $\simeq0.40$ and $h(B{:}\Bbar)$ to $\simeq0.64$---before saturating in turn.
The Unruh effect therefore does not create tripartite correlation here so much as
\emph{redistribute} the correlation already present,  moving it from the
$A{:}B$ channel into the channels involving the causally disconnected wedge.

Two important features are worth mentioning. First, $h(A{:}B)$ and
$h(A{:}\Bbar)$ \emph{merge} as $r\to\infty$ for all three states, the same convergence seen for the
reflected entropies themselves in Fig.~\ref{fig:saturation}: at infinite
acceleration Alice cannot tell the two wedges apart, so her correlation with
either is structurally identical. That this
pattern reproduces the fermionic one \cite{Basak2023} indicates that it reflects
the entanglement structure of the of the states themselves rather than the statistics of the
field. Second, the GHZ $h(A{:}B)$ is
\emph{non-monotonic}, peaking near $r\approx0.8$ at $\simeq0.29$ before
relaxing to its saturation floor. 


\section{From Rindler to Schwarzschild}
\label{sec:blackhole}
 
The construction so far of this paper is not tied to flat space. A bifurcate Killing
horizon is all that the two-mode squeezed structure of Sec.~\ref{sec:tmsv}
requires, and the Schwarzschild horizon supplies one, which we discuss in this section.
 
Consider a massless scalar field on the background of a Schwarzschild black hole of mass $M$, with
surface gravity $\kappa=1/4M$. Separating the field in the static exterior gives
Boulware modes labelled by $(\omega,\ell,m)$ with $\omega>0$; these are positive
frequency with respect to the Killing time $t$ and define the static vacuum, the
state a Boulware observer calls empty. As in the Rindler problem, the exterior is
only half the story: the maximally extended (Kruskal) geometry has a second
static region II behind the bifurcation surface, and each exterior mode
$b_{\omega\ell m}$ has a partner $b_{\omega\ell m}^{\rm II}$ supported there.
 
The Kruskal modes---those analytic across the future horizon, and therefore
regular for an infalling observer---are the natural ``in'' modes; the state they
annihilate is the Hartle--Hawking vacuum $\ket{0}_{\mathrm{HH}}$. Analytic continuation of
the Boulware modes to the Kruskal patch mixes each exterior mode with its region
II partner, giving a Bogoliubov transformation of precisely the two-mode squeezed
form used throughout this paper. For mode by mode this reads
\begin{equation}
\ket{0}_{\mathrm{HH}}\Big|_{\omega\ell m}=\frac{1}{\cosh r_\omega}\sum_{n=0}^{\infty}
\tanh^{n}\! r_\omega\,\ket{n}_{B}\ket{n}_{\Bbar}\;,
\label{eq:bhKruskal}
\end{equation}
which is eq.~\eqref{eq:vac} with $B$ the exterior mode and $\Bbar$ its partner.
The squeezing parameter is now fixed not by an acceleration but by the surface
gravity, which in units $G=\hbar=c=k_B=1$ reads
\begin{equation}
\tanh r_\omega=e^{-\pi\omega/\kappa}=e^{-\omega/2T_H},
\qquad T_H=\frac{\kappa}{2\pi}=\frac{1}{8\pi M} .
\label{eq:tanhBH}
\end{equation}
The single-excitation state $\ket{1}_{\mathrm{HH}}$ follows by acting with the Kruskal
creation operator exactly as in Sec.~\ref{sec:setup}, reproducing
eq.~\eqref{eq:one} with $r\to r_\omega$. Tracing out region II leaves the
exterior mode thermal which is given as
\begin{equation}
\nn_\omega=\sinh^2 r_\omega=\frac{1}{e^{\omega/T_H}-1},
\label{eq:nBH}
\end{equation}
the Planck spectrum at the Hawking temperature: eq.~\eqref{eq:tanhBH} is
equivalent to the statement $\tanh^2 r_\omega=e^{-\beta_H\omega}$ used in
Sec.~\ref{sec:tmsv}, so the Hartle--Hawking state is the thermofield double of
the exterior algebra at temperature $T_H$.
 
The whole dictionary is therefore given by the substitution
\begin{equation}
\begin{aligned}
a&\longrightarrow\kappa=\frac{1}{4M},\\
r&\longrightarrow r_\omega=\operatorname{arctanh}\big(e^{-\omega/2T_H}\big).
\end{aligned}
\label{eq:bhdict}
\end{equation}
That this must be so is guaranteed by the near-horizon geometry: a static
observer at areal radius $\rho$ has proper acceleration
$a_{\rm loc}=M/\big(\rho^2\sqrt{1-2M/\rho}\big)$ and measures the
Tolman-blueshifted temperature $T_{\rm loc}=T_H/\sqrt{1-2M/\rho}$, and these
satisfy $T_{\rm loc}\to a_{\rm loc}/2\pi$ as $\rho\to2M$. The flat-space results
of Secs.~\ref{sec:reflected}--\ref{sec:markov} are thus literally the
near-horizon limit of the black-hole problem, and the Unruh temperature of the
earlier sections is the locally measured Hawking temperature.
 
To pose the same question one must say who holds what, and the two roles are not
symmetric. Alice ($A$) remains at rest far from the black hole, where the geometry
is effectively flat and her mode is un-effected by the horizon; for the tripartite
states Charlie ($C$) does likewise. Because no Bogoliubov transformation acts on
their modes, the states of Sec.~\ref{sec:setup} populate only the lowest two
levels of each, and the two-dimensional span $\{\ket0,\ket1\}$ is invariant---so
$A$ and $C$ may be treated as qubits without approximation.
 
In contrast, the observer Bob ($B$), whose detector couples to a single Boulware mode $(\omega,\ell,m)$, stays  outside the black hole horizon in the exterior region. This is the black-hole counterpart of the uniformly accelerated observer, held at a fixed $\rho$ by an accelerating rocket rather than falling into the black hole, whose  entanglement
degradation  has previously been analyzed in \cite{MartinMartinezGarayLeon2010}. That mode is a bosonic field mode, and it is
\emph{not} a qubit: the Bogoliubov transformation \eqref{eq:bhKruskal} spreads
each Minkowski (Kruskal) excitation over the entire Fock tower of the exterior
mode, whose mean occupation \eqref{eq:nBH} grows without bound as
$\omega/T_H\to0$. Bob's Hilbert space is therefore infinite dimensional
throughout, exactly as in the flat-space analysis, and this is precisely what
distinguishes the bosonic problem from its fermionic counterpart, where Pauli
exclusion truncates the same mode to two levels. The partner $\Bbar$ is the mode
beyond the horizon, causally inaccessible to Bob and therefore traced out.
 
The identity of $\Bbar$ depends on which spacetime one has in consideration, and the two
cases carry different physics even though the algebra is identical. In the
\emph{eternal} two-sided black hole in the Hartle--Hawking state, $\Bbar$ lives in
the second asymptotic region behind the Einstein--Rosen bridge; the global state
is precisely the thermofield double of Sec.~\ref{sec:tmsv}, dual to two entangled
copies of the boundary theory \cite{Israel1976,Maldacena2003}, and
$\SR(B{:}\Bbar)$ is a correlation between the two observers in the two exterior regions. In a \emph{physical collapse} geometry the appropriate state is the Unruh
vacuum, which is thermal only in the \emph{outgoing} sector: the modes
analytic in the affine parameter on the past horizon are pairwise squeezed
with partners inside the black hole, with the same squeezing parameter
\eqref{eq:tanhBH}, while the ingoing sector remains in the ``in'' vacuum.
We therefore take Bob's detector to couple to an outgoing Boulware mode, for
which the two-mode structure \eqref{eq:bhKruskal} holds at late retarded
times; there is then no second exterior, the partner modes lie inside the
horizon, and tracing out $\Bbar$ means tracing over the black-hole interior.
Then $\SR(A{:}B)$ measures the correlation an external observer retains with
a Hawking quanta, and $\SR(B{:}\Bbar)$ the correlation between an emitted
quanta and its interior partner.
 
With the parties fixed, the accelerated Bell, GHZ, and W states of
Sec.~\ref{sec:setup} are constructed exactly as before, substituting
\eqref{eq:bhKruskal} and its excited partner for Bob's states.
 
Every step from Sec.~\ref{sec:reflected} onward uses the state only through the
single number $t=\tanh r$. Nothing in the derivations refers to the origin of
$t$. Consequently, at fixed $(\omega,\ell,m)$ and with $t=e^{-\omega/2T_H}$:
the three operators $\tau_B,\sigma_B,\kappa_B$ of eq.~\eqref{eq:threeops} retain
their form; the charge $\hat q=\hat n_B-\hat n_A$ is conserved for the same
reason (each Alice excitation is tied to one extra exterior quantum), so
$\rho_{AB}$ is again block diagonal in $2\times2$ sectors; the Bell sectors are
still rank one and the W sectors still rank two; and the Gram construction
still collapses to a $4\times4$ matrix with a single coherence. The semi-analytic
Lerch expressions \eqref{eq:bellclosed}, the GHZ affinity \eqref{eq:g} with its
closed form \eqref{eq:ghzSR} all hold verbatim mode by mode.

In particular the saturation floors are unchanged, and they are pure numbers,
independent of the black-hole mass, which are given as 
\begin{equation}
\begin{aligned}
\SR^{(\mathrm{GHZ})}(A{:}B)&\to H_2\!\Big(\tfrac{1+\sqrt\pi/2}{2}\Big)=0.315\ ,\\
\SR^{(\mathrm{B})}(A{:}B)&\to1.7572\ ,\\
\SR^{(\mathrm{W})}(A{:}B)&\to0.7523\ .
\end{aligned}
\label{eq:bhfloors}
\end{equation}
This is worth emphasising. The saturation floor is set by the affinity of the two Bob states
in the deep-thermal limit---a property of Bose statistics---and not by any
feature of the geometry. A solar-mass and a supermassive black hole degrade
correlations to the same residual value; only the frequency at which a given mode
sits on the curve depends on $M$. By the same token the inter-wedge divergence
$\SR(B{:}\Bbar)\sim4r_\omega\log_2e$ persists, now as a statement about the
correlation between soft exterior quanta and their partners.
 
The substantive change is what it means to move along the horizontal axis of
Figs.~\ref{fig:reflected}--\ref{fig:saturation}. In the Unruh problem one dials
the acceleration at fixed $\omega$. For a black hole $T_H$ is fixed by the mass,
and \eqref{eq:tanhBH} makes $r_\omega$ a function of the dimensionless ratio
$\omega/T_H$: a single black hole realises the entire curve at once, across its
spectrum. The soft modes $\omega\ll T_H$ sit at large $r_\omega$ and are maximally
degraded; while the hard modes $\omega\gg T_H$ have $r_\omega\to0$, so $\Bbar$ decouples,
$\rho_{AB}$ is pure, and $\SR(A{:}B)\to2H_2(\alpha^2)$, equal to $2$ at
$\alpha=1/\sqrt2$.
 
The two remaining pairings are read across the spectrum in the same way, and each
acquires a distinctly gravitational reading. For $A{:}\Bbar$ the flat-space
result was that it approaches $\SR(A{:}B)$ from below, the two meeting at the
common floor; in the black-hole variables this convergence is governed by
$\tanh r_\omega=e^{-\omega/2T_H}$, which tends to unity as $\omega/T_H\to0$. For
the GHZ state the statement is exact, since $g=\tanh r_\omega\,\tilde g$ by
\eqref{eq:ggtilde}. Physically, an exterior observer correlated with a soft
Hawking mode retains as much correlation with the partner mode behind the horizon
as with the one in front of it: in the deep infrared the two sides of the horizon
are statistically indistinguishable to Alice. For hard modes the situation
reverses, $r_\omega\to0$ and $\SR(A{:}\Bbar)\to0$, the partner decoupling
entirely. The saturation floors of \eqref{eq:bhfloors} are therefore approached by
\emph{both} Alice pairings, and are mass-independent for the same reason.
 
The inter-wedge quantity behaves differently, and is the one place where the
bosonic character of the field has an unbounded effect. Since
$\SR(B{:}\Bbar)\sim4r_\omega\log_2e$ and $r_\omega$ diverges logarithmically as
$\omega/T_H\to0$, the correlation between a mode and its partner grows without
bound in the infrared, at a rate set only by $T_H$. Its interpretation depends on
the spacetime, as discussed above: for the eternal black hole in
the Hartle--Hawking state it is a correlation between the two asymptotic regions,
and its growth is the statement that the thermofield double becomes ever more
strongly entangled for soft modes; in a collapse geometry it is the correlation
between an emitted Hawking quantum and its interior partner, so its divergence is
the reflected-entropy counterpart of the familiar infrared divergence of horizon
entanglement entropy. This is also the pairing that fixes the scope of
Table~\ref{tab:bhvalues}: because the divergence is mode by mode, any sum over
$(\omega,\ell,m)$ inherits an infrared sensitivity that the bounded Alice
pairings do not, and it has no fermionic analogue, Pauli exclusion setting the
corresponding quantity at unit value.
 
Table~\ref{tab:bhvalues} makes this concrete for the three states.
\begin{table*}[t]
\centering
\begin{tabular}{c c c c c c}
\hline\hline
$\omega/T_H$ & $t=\tanh r_\omega$ & $\nn_\omega$ &
$\SR^{(\mathrm{B})}$ & $\SR^{(\mathrm{W})}$ & $\SR^{(\mathrm{GHZ})}$\\
\hline
$\to 0$ & $\to 1$ & $\to \infty$ & $1.757$ & $0.752$ & $0.315$\\
\hline
$0.1$ & $0.951$ & $9.51$   & $1.780$ & $0.779$ & $0.369$\\
$0.5$ & $0.779$ & $1.542$  & $1.853$ & $0.883$ & $0.567$\\
$1$   & $0.607$ & $0.582$  & $1.911$ & $1.001$ & $0.736$\\
$2$   & $0.368$ & $0.157$  & $1.967$ & $1.185$ & $0.903$\\
$5$   & $0.082$ & $0.007$  & $1.998$ & $1.422$ & $0.995$\\
$10$  & $0.007$ & $0.00005$& $2.000$ & $1.483$ & $1.000$\\
\hline\hline
\end{tabular}
\caption{Reflected entropy $\SR(A{:}B)$ across the Hawking spectrum, at
$\alpha=1/\sqrt2$ for Bell and GHZ and $\alpha=1/\sqrt3$ for the W state. The first row is the infinite-squeezing saturation floor
\eqref{eq:bhfloors}. Hard modes ($\omega\gg T_H$) are essentially unaffected by
the horizon; soft modes ($\omega\ll T_H$) approach the saturation floor.}
\label{tab:bhvalues}
\end{table*}
Two features are worth mentioning. The approach to the saturation floor is slow: even at
$\omega=0.1\,T_H$, the reflected entropy for Bell state is $1.780$, which exceeds the floor value of $1.757$, and one
must go to $\omega\lesssim0.01\,T_H$ to reach close to the saturation value. Conversely the
degradation is already negligible by $\omega\simeq5\,T_H$. The loss of
correlation is confined to the infrared tail (Table~\ref{tab:bhvalues}), which carries little energy but
dominates the mode count.
 
The W state deserves separate comment, since it behaves differently from the
other two at both ends of the spectrum. In the hard-mode limit $\omega\gg T_H$ it
does \emph{not} approach the value $2H_2(\alpha^2)$ characteristic of a pure
$\rho_{AB}$, but the smaller value $1.488$. The reason is structural rather than
gravitational: tracing out Charlie from a W state leaves $\rho_{AB}$ mixed
\emph{already in the inertial limit}, because the three excitations are shared
symmetrically and Charlie's branch is not perfectly correlated with any single
Alice--Bob configuration. The W state therefore begins its degradation from an
intermediate value, between the $2$ of the Bell state (whose $\rho_{AB}$ is pure
at $r=0$) and the $1$ of GHZ (whose $\rho_{AB}$ is classically correlated).

These values are per mode, and they refer to a detector coupled to a single
Boulware mode inside the potential barrier. An observer at infinity receives
the mode filtered by the greybody transmissivity $\Gamma_{\omega\ell m}$,
which acts as a Gaussian attenuation channel on $B$: it preserves the charge
grading and hence the sector structure, but the sectors are no longer rank
one, so the closed forms of Sec.~\ref{sec:bellanalytic} do not carry over
and the values at infinity will generically differ from those of
Table~\ref{tab:bhvalues}. We note that no monotonicity argument fixes the
direction of the change a priori: the reflected entropy is not monotonic
under local channels \cite{HaydenLemmSorce2023}, so the effect of
$\Gamma_{\omega\ell m}$ must be computed rather than bounded. Since
attenuation preserves the block decomposition, this is a straightforward
extension of the sector method, which we leave for future work. Any total
quantity further requires the sum over $(\omega,\ell,m)$, which for
$\SR(B{:}\Bbar)$ inherits an infrared sensitivity from the divergence at
small $\omega$.

The behavior of the Markov gap in the Schwarzschild geometry is the same as
that found in the Rindler case of Sec.~\ref{sec:markov}: both problems depend
on the state only through $t$. The Markov gaps of the three states are
therefore those of Fig.~\ref{fig:markov}, with the horizontal axis relabelled
by $r\to r_\omega=\operatorname{arctanh}\big(e^{-\omega/2T_H}\big)$. Since
$r_\omega$ is a monotonically decreasing function of the ratio $\omega/T_H$
alone, the relabelling can be read in two ways. At fixed $T_H$---a single
black hole of given mass---what was a statement about increasing acceleration
becomes one about decreasing frequency: moving toward the deep infrared of
the Hawking spectrum plays the role of the infinite-acceleration limit, and a
single black hole realises the entire family of curves across its spectrum.
At fixed $\omega$, by contrast, traversing the curves requires increasing
$T_H$, i.e.\ decreasing the mass; for a single black hole this is realised
dynamically by evaporation. In the quasi-static regime, where the mass
changes slowly on the timescale set by the mode frequency, a detector coupled
to a fixed Boulware frequency then sweeps along the curves of
Fig.~\ref{fig:markov} as the black hole shrinks, its Markov gap flowing toward the
saturation plateau. In particular, for the initially bipartite Bell and GHZ
states, whose Markov gap vanishes in the inertial limit, evaporation progressively
generates tripartite correlation among $A$, $B$ and the interior
partner; for the W state it instead redistributes the tripartite correlation
already present into the channels involving the interior. A treatment beyond
the quasi-static approximation, including the mode mixing induced by the
time-dependent horizon, lies outside the single-mode framework used here.

 \section{Summary and discussion}
\label{sec:discussion}

In this work, we have investigated the redistribution of bosonic field correlations in non-inertial frames and black hole spacetimes using the reflected entropy and the Markov gap. By moving beyond standard entanglement measures, which vanish on
classically correlated states and therefore misreport the correlations of
generically mixed states, we demonstrated that the degradation of bosonic correlations under the Unruh effect is structurally richer than previously understood. 

The central technical hurdle in the continuous-variable regime---the unbounded occupation of bosonic Rindler modes---was resolved by identifying a conserved $U(1)$ charge. This conservation law enabled the exact decomposition of the infinite-dimensional reduced density matrices into an infinite family of two-dimensional sectors. Consequently, we obtained semi-analytic and exact closed-form expressions for the reflected entropy of the Bell, GHZ, and W states across the full range of acceleration. 

Our primary physical conclusion is that the reflected entropy between the inertial and accelerated observers (the Alice--Bob pair) does not vanish at infinite acceleration. Instead, it degrades and subsequently saturates at a finite, non-zero, state-dependent floor. This saturation confirms that while bipartite entanglement is diminished
by the Unruh thermal bath, a resilient core of correlation---classical rather than
entanglement---survives; the same conclusion is reached independently by the
mutual information, which is a genuine correlation monotone and likewise saturates
at a nonzero value. Furthermore, we established that the defining signature of bosonic statistics lies not in the degradation of the Alice--Bob pair, but in the unbounded growth of the inter-wedge (Bob--anti-Bob) pair. Driven by the unbounded thermal excitation of the vacuum, the inter-wedge reflected entropy diverges linearly with the squeezing parameter, standing in sharp contrast to the bounded behavior characteristic of fermionic systems. 

The Markov gap analysis further refined this picture by revealing that the residual correlations retain a genuinely tripartite structure. For initially bipartite states, the Unruh effect actively generates tripartite correlations among the accessible and inaccessible wedges that remain invisible to bipartite measures and local recovery channels. 

Crucially, because these results depend on the field state entirely through the Bogoliubov squeezing parameter, the mathematical structure transfers intact to the Schwarzschild black hole scenario. The infinite-acceleration saturation floors of the Unruh problem manifest as mass-independent constants across the Hawking spectrum, placing a strict lower bound on the correlation an external observer
retains with a soft Hawking quantum.

As for the future direction, the most immediate extension is the rotating black holes case, where superradiant modes exhibit a negative Klein--Gordon norm in the ergoregion. Because superradiant amplification is an intrinsically bosonic phenomenon governed by a re-paired Bogoliubov transformation, evaluating the reflected entropy in the Kerr geometry promises to uncover novel correlation structures without any fermionic counterpart. Finally, extending this exact sector-decomposition method to account for greybody factors or relaxing the single-mode approximation would provide a more operational description of detector-field interactions outside a horizon.

\section*{Acknowledgements} \noindent
I would like to thank Ayan Mukhopadhyay for his hospitality at the Pontificia Universidad Católica de Valparaíso, Curauma.  This work is supported by the ANID FONDECYT Postdoctorado
grant number 3240055.

\appendix

\section{Explicit reduced density matrices}
\label{app:rho}

All reduced density matrices are built from three operators on $\mathcal{H}_B$
($t\equiv\tanh r$), which are
\begin{align}
\tau_B &=(1-t^2)\sum_{n\ge0}t^{2n}\ketbra{n}{n},\\
\sigma_B &=(1-t^2)^2\sum_{n\ge0}(n+1)\,t^{2n}\ketbra{n+1}{n+1},\\
\kappa_B &=(1-t^2)^{3/2}\sum_{n\ge0}\sqrt{n+1}\,t^{2n}\ketbra{n}{n+1},
\end{align}
with $\tr\tau_B=\tr\sigma_B=1$ ($\tau_B$ is the thermal reduced state of the
two-mode squeezed vacuum, $\sigma_B$ its single-excitation partner, $\kappa_B$ the
coherence from Alice's superposition). In the qubit-$A$ basis
$\{\ket0,\ket1\}$ 
\begin{align}
\rho^{(B)}_{AB}&=
\begin{pmatrix}\alpha^2\tau_B & \alpha\sqrt{1-\alpha^2}\,\kappa_B\\
\alpha\sqrt{1-\alpha^2}\,\kappa_B^{\dagger} & (1-\alpha^2)\sigma_B\end{pmatrix},
\label{eq:rhoABbellApp}\\
\rho^{(\mathrm{GHZ})}_{AB}&=
\begin{pmatrix}\alpha^2\tau_B & 0\\ 0 & (1-\alpha^2)\sigma_B\end{pmatrix},
\label{eq:rhoABghzApp}\\
\rho^{(W)}_{AB}&=
\begin{pmatrix}\alpha^2\tau_B+(1-2\alpha^2)\sigma_B & \alpha\sqrt{1-2\alpha^2}\,\kappa_B^{\dagger}\\
\alpha\sqrt{1-2\alpha^2}\,\kappa_B & \alpha^2\tau_B\end{pmatrix}.
\label{eq:rhoABwApp}
\end{align}
The $A{:}\Bbar$ matrices have the same form with $\sigma_B$ replaced by its
unshifted partner $\tilde\sigma_B=(1-t^2)^2\sum_n(n+1)t^{2n}\ketbra{n}{n}$ and
$\kappa_B$ by $t\,\tilde\kappa_B$, $\tilde\kappa_B=(1-t^2)^{3/2}\sum_n\sqrt{n+1}
t^{2n}\ketbra{n+1}{n}$. Tracing out $A$ gives the rank-two inter-wedge state
\begin{equation}
\rho_{B\Bbar}=p_0\ketbra{\psi_0}{\psi_0}+p_1\ketbra{\psi_1}{\psi_1},
\end{equation}
with $\ket{\psi_0}=\sqrt{1-t^2}\sum_n t^n\ket{n}_B\ket{n}_{\Bbar}$,
$\ket{\psi_1}=(1-t^2)\sum_n\sqrt{n+1}\,t^n\ket{n{+}1}_B\ket{n}_{\Bbar}$, and
$(p_0,p_1)=(\alpha^2,1-\alpha^2)$ for Bell/GHZ, $(2\alpha^2,1-2\alpha^2)$ for W.
Equation~\eqref{eq:rhoBBstar} follows from purifying this and tracing the mirror
wedge; since $\ketbra{\psi_0}{\psi_0}$ and $\ketbra{\psi_1}{\psi_1}$ are
orthogonal, $S_{B\Bbar}=H_2(p_0,p_1)$ is independent of $r$, while $S_B,S_{\Bbar}$
grow like $2r$, forcing $\SR(B{:}\Bbar)$ to diverge.

\bibliographystyle{apsrev4-1}
\bibliography{references}

@article{Davies1975,
doi = {10.1088/0305-4470/8/4/022},
url = {https://doi.org/10.1088/0305-4470/8/4/022},
year = {1975},
month = {apr},
publisher = {},
volume = {8},
number = {4},
pages = {609},
author = {P C W Davies},
title = {Scalar production in Schwarzschild and Rindler metrics},
journal = {Journal of Physics A: Mathematical and General}
}

@article{Pan:2008yr,
    author = "Pan, Qiyuan and Jing, Jiliang",
    title = "{Hawking radiation, Entanglement and Teleportation in background of an asymptotically flat static black hole}",
    eprint = "0809.0811",
    archivePrefix = "arXiv",
    primaryClass = "gr-qc",
    doi = "10.1103/PhysRevD.78.065015",
    journal = "Phys. Rev. D",
    volume = "78",
    pages = "065015",
    year = "2008"
}

@article{Teng:2026dyr,
    author = "Teng, Xiao-Wei and Xu, Rui-Yang and Yang, Hui-Chen and Wu, Shu-Min",
    title = "{Bosonic and fermionic mutual information of N-partite systems in dilaton black hole background}",
    eprint = "2603.18439",
    archivePrefix = "arXiv",
    primaryClass = "gr-qc",
    doi = "10.1016/j.nuclphysb.2026.117559",
    journal = "Nucl. Phys. B",
    volume = "1029",
    pages = "117559",
    year = "2026"
}

@article{Alsing:2012wf,
    author = "Alsing, Paul M. and Fuentes, Ivette",
    title = "{Observer dependent entanglement}",
    eprint = "1210.2223",
    archivePrefix = "arXiv",
    primaryClass = "quant-ph",
    doi = "10.1088/0264-9381/29/22/224001",
    journal = "Class. Quant. Grav.",
    volume = "29",
    pages = "224001",
    year = "2012"
}

@article{Unruh1976,
  author       = {Unruh, W. G.},
  title        = {Notes on black-hole evaporation},
  journal      = {Phys. Rev. D},
  volume       = {14},
  pages        = {870},
  year         = {1976},
  doi          = {10.1103/PhysRevD.14.870}
}

@article{FuentesSchuller2005,
  author       = {Fuentes-Schuller, I. and Mann, R. B.},
  title        = {Alice falls into a black hole: Entanglement in non-inertial frames},
  journal      = {Phys. Rev. Lett.},
  volume       = {95},
  pages        = {120404},
  year         = {2005},
  doi          = {10.1103/PhysRevLett.95.120404}}

@article{Alsing2006,
  author       = {Alsing, P. M. and Fuentes-Schuller, I. and Mann, R. B. and Tessier, T. E.},
  title        = {Entanglement of {Dirac} fields in non-inertial frames},
  journal      = {Phys. Rev. A},
  volume       = {74},
  pages        = {032326},
  year         = {2006},
  eprint       = {quant-ph/0603269},
  archivePrefix= {arXiv},
  doi          = {10.1103/PhysRevA.74.032326}
}

@article{Adesso2007,
  author       = {Adesso, G. and Fuentes-Schuller, I. and Ericsson, M.},
  title        = {Continuous-variable entanglement sharing in noninertial frames},
  journal      = {Phys. Rev. A},
  volume       = {76},
  pages        = {062112},
  year         = {2007},
  eprint       = {quant-ph/0701074},
  archivePrefix= {arXiv},
  doi          = {10.1103/PhysRevA.76.062112}
}

@article{NasrEsfahani2011,
  author       = {{Nasr Esfahani}, B. and Shamirzaie, M. and Soltani, M.},
  title        = {Tripartite entanglements in non-inertial frames},
  journal      = {Phys. Rev. D},
  volume       = {84},
  pages        = {025024},
  year         = {2011},
  eprint       = {1103.0258},
  archivePrefix= {arXiv},
  doi          = {10.1103/PhysRevD.84.025024}
}

@article{Hwang2011,
  author       = {Hwang, M.-R. and Park, D. and Jung, E.},
  title        = {Tripartite entanglement in a noninertial frame},
  journal      = {Phys. Rev. A},
  volume       = {83},
  pages        = {012111},
  year         = {2011},
  doi          = {10.1103/PhysRevA.83.012111}
}

@article{TianJing2013,
  author       = {Tian, Z. and Jing, J.},
  title        = {Measurement-induced-nonlocality via the {Unruh} effect},
  journal      = {Annals Phys.},
  volume       = {333},
  pages        = {76},
  year         = {2013},
  eprint       = {1301.5981},
  archivePrefix= {arXiv},
  doi          = {10.1016/j.aop.2013.02.012}
}

@article{Datta2009,
  author       = {Datta, A.},
  title        = {Quantum discord between relatively accelerated observers},
  journal      = {Phys. Rev. A},
  volume       = {80},
  pages        = {052304},
  year         = {2009},
  eprint       = {0905.3301},
  archivePrefix= {arXiv},
  doi          = {10.1103/PhysRevA.80.052304}
}

@article{WangDengJing2010,
  author       = {Wang, J. and Deng, J. and Jing, J.},
  title        = {Classical correlation and quantum discord sharing of {Dirac} fields in noninertial frames},
  journal      = {Phys. Rev. A},
  volume       = {81},
  pages        = {052120},
  year         = {2010},
  eprint       = {0912.4129},
  archivePrefix= {arXiv},
  doi          = {10.1103/PhysRevA.81.052120}
}

@article{MartinMartinezLeon2010,
  author       = {Mart{\'i}n-Mart{\'i}nez, E. and Le{\'o}n, J.},
  title        = {Quantum correlations through event horizons: {Fermionic} versus bosonic entanglement},
  journal      = {Phys. Rev. A},
  volume       = {81},
  pages        = {032320},
  year         = {2010},
  eprint       = {1001.4302},
  archivePrefix= {arXiv},
  doi          = {10.1103/PhysRevA.81.032320}
}

@article{TakayanagiUmemoto2017,
  author       = {Takayanagi, T. and Umemoto, K.},
  title        = {Entanglement of purification through holographic duality},
  journal      = {Nature Phys.},
  volume       = {14},
  pages        = {573},
  year         = {2018},
  eprint       = {1708.09393},
  archivePrefix= {arXiv},
  doi          = {10.1038/s41567-018-0075-2}
}

@article{Nguyen2017,
  author       = {Nguyen, P. and Devakul, T. and Halbasch, M. G. and Zaletel, M. P. and Swingle, B.},
  title        = {Entanglement of purification: from spin chains to holography},
  journal      = {JHEP},
  volume       = {01},
  pages        = {098},
  year         = {2018},
  eprint       = {1709.07424},
  archivePrefix= {arXiv},
  doi          = {10.1007/JHEP01(2018)098}
}

@article{Bruschi2010,
  author       = {Bruschi, D. E. and Louko, J. and Mart{\'i}n-Mart{\'i}nez, E. and Dragan, A. and Fuentes, I.},
  title        = {Unruh effect in quantum information beyond the single-mode approximation},
  journal      = {Phys. Rev. A},
  volume       = {82},
  pages        = {042332},
  year         = {2010},
  eprint       = {1007.4670},
  archivePrefix= {arXiv},
  doi          = {10.1103/PhysRevA.82.042332}
}

@article{DuttaFaulkner2021,
  author       = {Dutta, S. and Faulkner, T.},
  title        = {A canonical purification for the entanglement wedge cross-section},
  journal      = {JHEP},
  volume       = {3},
  pages        = {178},
  year         = {2021},
  eprint       = {1905.00577},
  archivePrefix= {arXiv},
  doi          = {10.1007/JHEP03(2021)178}
}

@article{HaydenParrikarSorce2021,
  author       = {Hayden, P. and Parrikar, O. and Sorce, J.},
  title        = {The {Markov} gap for geometric reflected entropy},
  journal      = {JHEP},
  volume       = {10},
  pages        = {047},
  year         = {2021},
  eprint       = {2107.00009},
  archivePrefix= {arXiv},
  doi          = {10.1007/JHEP10(2021)047}
}

@article{HaydenLemmSorce2023,
  author       = {Hayden, P. and Lemm, M. and Sorce, J.},
  title        = {Reflected entropy is not a correlation measure},
  journal      = {Phys. Rev. A},
  volume       = {107},
  pages        = {L050401},
  year         = {2023},
  eprint       = {2302.10208},
  archivePrefix= {arXiv},
  doi          = {10.1103/PhysRevA.107.L050401}
}

@article{FawziRenner2015,
  author       = {Fawzi, O. and Renner, R.},
  title        = {Quantum conditional mutual information and approximate {Markov} chains},
  journal      = {Commun. Math. Phys.},
  volume       = {340},
  pages        = {575},
  year         = {2015},
  eprint       = {1410.0664},
  archivePrefix= {arXiv},
  doi          = {10.1007/s00220-015-2466-x}
}

@article{Coffman2000,
  author       = {Coffman, V. and Kundu, J. and Wootters, W. K.},
  title        = {Distributed entanglement},
  journal      = {Phys. Rev. A},
  volume       = {61},
  pages        = {052306},
  year         = {2000},
  eprint       = {quant-ph/9907047},
  archivePrefix= {arXiv},
  doi          = {10.1103/PhysRevA.61.052306}
}

@article{Basak2023,
  author       = {Basak, J. K. and Giataganas, D. and Mondal, S. and Wen, W.-Y.},
  title        = {Reflected entropy and {Markov} gap in non-inertial frames},
  journal      = {Phys. Rev. D},
  volume       = {108},
  pages        = {125009},
  year         = {2023},
  eprint       = {2306.17490},
  archivePrefix= {arXiv},
  doi          = {10.1103/PhysRevD.108.125009}
}

@article{CrispinoHiguchiMatsas2008,
  author       = {Crispino, L. C. B. and Higuchi, A. and Matsas, G. E. A.},
  title        = {The {Unruh} effect and its applications},
  journal      = {Rev. Mod. Phys.},
  volume       = {80},
  pages        = {787},
  year         = {2008},
  eprint       = {0710.5373},
  archivePrefix= {arXiv},
  doi          = {10.1103/RevModPhys.80.787}
}

@article{TakahashiUmezawa1975,
  author       = {Takahashi, Y. and Umezawa, H.},
  title        = {Thermo field dynamics},
  journal      = {Int. J. Mod. Phys. B},
  volume       = {10},
  pages        = {1755},
  year         = {1996},
  note         = {reprint of Collect. Phenom. \textbf{2}, 55 (1975)},
  doi          = {10.1142/S0217979296000817}
}

@article{Israel1976,
  author       = {Israel, W.},
  title        = {Thermo-field dynamics of black holes},
  journal      = {Phys. Lett. A},
  volume       = {57},
  pages        = {107},
  year         = {1976},
  doi          = {10.1016/0375-9601(76)90178-X}
}

@article{BisognanoWichmann1975,
  author       = {Bisognano, J. J. and Wichmann, E. H.},
  title        = {On the duality condition for a {Hermitian} scalar field},
  journal      = {J. Math. Phys.},
  volume       = {16},
  pages        = {985},
  year         = {1975},
  doi          = {10.1063/1.522605}
}

@article{Sewell1982,
  author       = {Sewell, G. L.},
  title        = {Quantum fields on manifolds: {PCT} and gravitationally induced thermal states},
  journal      = {Annals Phys.},
  volume       = {141},
  pages        = {201},
  year         = {1982},
  doi          = {10.1016/0003-4916(82)90285-8}
}

@article{Maldacena2003,
  author       = {Maldacena, J. M.},
  title        = {Eternal black holes in anti-de {Sitter}},
  journal      = {JHEP},
  volume       = {04},
  pages        = {021},
  year         = {2003},
  eprint       = {hep-th/0106112},
  archivePrefix= {arXiv},
  doi          = {10.1088/1126-6708/2003/04/021}
}

@article{BuenoCasini2020scalars,
  author       = {Bueno, P. and Casini, H.},
  title        = {Reflected entropy for free scalars},
  journal      = {JHEP},
  volume       = {11},
  pages        = {148},
  year         = {2020},
  eprint       = {2008.11373},
  archivePrefix= {arXiv},
  doi          = {10.1007/JHEP11(2020)148}
}

@article{BuenoCasini2020fermions,
  author       = {Bueno, P. and Casini, H.},
  title        = {Reflected entropy, symmetries and free fermions},
  journal      = {JHEP},
  volume       = {05},
  pages        = {103},
  year         = {2020},
  eprint       = {2003.09546},
  archivePrefix= {arXiv},
  doi          = {10.1007/JHEP05(2020)103}
}

@article{AkersRath2020,
  author       = {Akers, C. and Rath, P.},
  title        = {Entanglement wedge cross sections require tripartite entanglement},
  journal      = {JHEP},
  volume       = {04},
  pages        = {208},
  year         = {2020},
  eprint       = {1911.07852},
  archivePrefix= {arXiv},
  doi          = {10.1007/JHEP04(2020)208}
}

@article{Camargo2021,
  author       = {Camargo, H. A. and Hackl, L. and Heller, M. P. and Jahn, A. and Windt, B.},
  title        = {Long distance entanglement of purification and reflected entropy in conformal field theory},
  journal      = {Phys. Rev. Lett.},
  volume       = {127},
  pages        = {141604},
  year         = {2021},
  eprint       = {2102.00013},
  archivePrefix= {arXiv},
  doi          = {10.1103/PhysRevLett.127.141604}
}

@article{DuttaFaulknerLin2023,
  author       = {Dutta, S. and Faulkner, T. and Lin, S.},
  title        = {The reflected entanglement spectrum for free fermions},
  journal      = {JHEP},
  volume       = {02},
  pages        = {223},
  year         = {2023},
  eprint       = {2211.17255},
  archivePrefix= {arXiv},
  doi          = {10.1007/JHEP02(2023)223}
}

@article{BerthiereParez2023,
  author       = {Berthiere, C. and Parez, G.},
  title        = {Reflected entropy and computable cross-norm negativity: Free theories and symmetry resolution},
  journal      = {Phys. Rev. D},
  volume       = {108},
  pages        = {054508},
  year         = {2023},
  eprint       = {2307.11009},
  archivePrefix= {arXiv},
  doi          = {10.1103/PhysRevD.108.054508}
}

@article{BerthiereChenChen2023,
  author       = {Berthiere, C. and Chen, B. and Chen, H.},
  title        = {Reflected entropy and {Markov} gap in {Lifshitz} theories},
  journal      = {JHEP},
  volume       = {09},
  pages        = {160},
  year         = {2023},
  eprint       = {2307.12247},
  archivePrefix= {arXiv},
  doi          = {10.1007/JHEP09(2023)160}
}

@article{AkersFaulknerLinRath2022,
  author       = {Akers, C. and Faulkner, T. and Lin, S. and Rath, P.},
  title        = {Reflected entropy in random tensor networks},
  journal      = {JHEP},
  volume       = {05},
  pages        = {162},
  year         = {2022},
  eprint       = {2112.09122},
  archivePrefix= {arXiv},
  doi          = {10.1007/JHEP05(2022)162}
}

@article{AkersPageCurve2022,
  author       = {Akers, C. and Faulkner, T. and Lin, S. and Rath, P.},
  title        = {The {Page} curve for reflected entropy},
  journal      = {JHEP},
  volume       = {06},
  pages        = {089},
  year         = {2022},
  eprint       = {2201.11730},
  archivePrefix= {arXiv},
  doi          = {10.1007/JHEP06(2022)089}
}

@article{LiChuZhou2020,
  author       = {Li, T. and Chu, J. and Zhou, Y.},
  title        = {Reflected entropy for an evaporating black hole},
  journal      = {JHEP},
  volume       = {11},
  pages        = {155},
  year         = {2020},
  eprint       = {2006.10846},
  archivePrefix= {arXiv},
  doi          = {10.1007/JHEP11(2020)155}
}

@article{LuLin2023,
  author       = {Lu, Y. and Lin, J.},
  title        = {The {Markov} gap in the presence of islands},
  journal      = {JHEP},
  volume       = {03},
  pages        = {043},
  year         = {2023},
  eprint       = {2211.06886},
  archivePrefix= {arXiv},
  doi          = {10.1007/JHEP03(2023)043}
}

@article{MartinMartinezLeon2010b,
  author       = {Mart{\'i}n-Mart{\'i}nez, E. and Le{\'o}n, J.},
  title        = {Population bound effects on bosonic correlations in non-inertial frames},
  journal      = {Phys. Rev. A},
  volume       = {81},
  pages        = {052305},
  year         = {2010},
  eprint       = {1003.3550},
  archivePrefix= {arXiv},
  doi          = {10.1103/PhysRevA.81.052305}
}

@article{MartinMartinezGarayLeon2010,
  author       = {Mart{\'i}n-Mart{\'i}nez, E. and Garay, L. J. and Le{\'o}n, J.},
  title        = {Unveiling quantum entanglement degradation near a {Schwarzschild} black hole},
  journal      = {Phys. Rev. D},
  volume       = {82},
  pages        = {064006},
  year         = {2010},
  eprint       = {1006.1394},
  archivePrefix= {arXiv},
  doi          = {10.1103/PhysRevD.82.064006}
}

@article{RichterOmar2015,
  author       = {Richter, B. and Omar, Y.},
  title        = {Degradation of entanglement between two accelerated parties: {Bell} states under the {Unruh} effect},
  journal      = {Phys. Rev. A},
  volume       = {92},
  pages        = {022334},
  year         = {2015},
  doi          = {10.1103/PhysRevA.92.022334}
}

@article{TorresArenas2019,
  author       = {Torres-Arenas, A. J. and Dong, Q. and Sun, G.-H. and Qiang, W.-C. and Dong, S.-H.},
  title        = {Entanglement measures of {W}-state in noninertial frames},
  journal      = {Phys. Lett. B},
  volume       = {789},
  pages        = {93},
  year         = {2019}
}

@article{WuZengCao2021,
  author       = {Wu, S.-M. and Zeng, H.-S. and Cao, H.-M.},
  title        = {Quantum coherence and distribution of {N}-partite bosonic fields in noninertial frame},
  journal      = {Class. Quantum Grav.},
  volume       = {38},
  pages        = {185007},
  year         = {2021},
  eprint       = {2201.00986},
  archivePrefix= {arXiv},
  doi          = {10.1088/1361-6382/ac1b09}
}

@article{LiWu2024,
  author       = {Li, W.-M. and Wu, S.-M.},
  title        = {Bosonic and fermionic coherence of {N}-partite states in the background of a dilaton black hole},
  journal      = {JHEP},
  volume       = {09},
  pages        = {144},
  year         = {2024},
  eprint       = {2407.07688},
  archivePrefix= {arXiv},
  doi          = {10.1007/JHEP09(2024)144}
}

\end{document}